\documentclass[twocolumns]{aa}

\usepackage{lineno,hyperref}

\usepackage{natbib}
\usepackage{amssymb}
\usepackage{subfig}
\usepackage{graphicx}

\usepackage{amsmath}    
\usepackage[dvipsnames]{xcolor} 
\usepackage[T1]{fontenc}
\usepackage{ae,aecompl}
\usepackage[english]{babel}

\usepackage{hyperref}
\usepackage{url}
\usepackage{soul}

\usepackage{graphicx}   
\usepackage{amsmath}    
\usepackage{amssymb}    
\usepackage{adjustbox}
\usepackage{threeparttable} 
\usepackage[normalem]{ulem}

\newcommand{\be}{\begin{equation}}
\newcommand{\ee}{\end{equation}}
\newcommand{\bea}{\begin{eqnarray}}

\newcommand{\eea}{\end{eqnarray}}
\DeclareMathAlphabet\mathbfcal{OMS}{cmsy}{b}{n}

\newcommand{\s}{\scriptscriptstyle}

\newcommand{\PC}{PC }

\hypersetup{colorlinks=true, citecolor=blue, linkcolor=blue}

\modulolinenumbers[5]

\begin{document}

\title{Past and present dynamics of the circumbinary moons in the Pluto-Charon system}

   \author{Cristian A. Giuppone
          \inst{1}\fnmsep\thanks{E-mail: cristian.giuppone@unc.edu.ar},
          Adrián Rodríguez\inst{2},
          Tatiana A. Michtchenko\inst{3},
          Amaury A. {de Almeida}\inst{3}
          }

   \institute{Universidad Nacional de C\'ordoba, Observatorio Astron\'omico - IATE. Laprida 854, 5000 C\'ordoba, Argentina
        \and
        Observatório do Valongo, Universidade Federal do Rio de Janeiro, Ladeira do Pedro Antônio 43, 20080-090, Rio de Janeiro, Brazil
        \and
        Instituto de Astronomia, Geof\'isica e Ci\^encias Atmosf\'ericas, USP, Rua do Mat\~ao 1226, S\~ao Paulo, SP 05508-090, Brazil
         }

\date{}

\abstract
{The Pluto-Charon (\PC) pair is usually thought of as a binary in the dual synchronous state, which is the endpoint of its tidal evolution. The discovery of the small circumbinary moons, Styx, Nix, Kerberos, and Hydra, placed close to the mean motions resonances (MMRs) 3/1, 4/1, 5/1, and 6/1 with Charon, respectively, reveals a complex dynamical architecture of the system. Several formation mechanisms for the \PC system have been proposed.}
{Assuming the hypothesis of the \textit{in-situ} formation of the moons, our goal is to analyse the past and current orbital dynamics of the satellite system. We plan to elucidate in which scenario the small moons can survive the rapid tidal expansion of the \PC binary.}
 {We study the past and current dynamics of the PC system through a large set of numerical integrations of the exact equations of motion, accounting for the gravitational interactions of the PC binary with the small moons and the tidal evolution, modelled by the constant time lag approach. We construct the stability maps in a pseudo-Jacobian coordinate system. In addition, considering a more realistic model, which accounts for the zonal harmonic $J_2$ of the Pluto's oblateness and the \textit{ad-hoc} accreting mass of Charon, we investigate the tidal evolution of the whole system.} 
{Our results show that, in the chosen reference frame, the current orbits of all satellites are nearly circular, nearly planar and nearly resonant with Charon that can be seen as an indicator of the convergent dissipative migration experimented by the system in the past. We verify that, under the assumption that Charon completes its formation during the tidal expansion, the moons can safely cross the main MMRs, without their motions being strongly excited and consequently ejected.}
{In the more realistic scenario proposed here, the small moons survive the tidal expansion of the \PC binary, without having to invoke the hypothesis of the resonant transport. Our results point out that the possibility to find additional small moons in the \PC system cannot be ruled out.}

\keywords{celestial mechanics -- Planets and satellites: dynamical evolution and stability-- methods: numerical -- planets and satellites: individual: Pluto}

\titlerunning{Past and present dynamics of the circumbinary moons in the Pluto-Charon system}
\authorrunning{Giuppone et al.}
\maketitle


\section{Introduction}
\label{sec:intro}

The Pluto-Charon (\PC)  binary has the mass ratio of {$\sim 0.122$} and is currently found in a dual synchronous state, which is the typical endpoint of the tidal evolution, over 1-10\,Myr for this particular system \citep[e.g.,][]{Farinella+1979, Correia+2020}. The orbit of Charon is almost circular, as confirmed by a 1$\sigma$ upper limit of $7.5 \times 10^{-5}$ \citep{Buie.2012}.
In the last 15 years, the system gained attention due to the discovery of the four small moons (Styx, Nix, Kerberos and Hydra) and their complex circumbinary configurations. Several formation mechanisms for that satellite system have been proposed \citep[][and references therein]{Kenyon+2021}. However, {none of the proposed scenarios of the past evolution of the whole system has provided strong conclusions} of whether the four small moons could survive the period of the tidal expansion of the \PC binary.

\begin{table*}
\begin{center}
\caption{Orbital elements and masses of the Pluto's satellites ; $R_{\s P}=1188$\,km and $m_{\s P}=1.303 \times 10^{22}$\,kg.}
\label{tab.Pluto}
\label{tab-init}
\resizebox{0.7\textwidth}{!}{%
\begin{tabular}{
|l|c|c|c|c|c|}
\hline
Parameter & Charon & Styx & Nix & Kerberos & Hydra\\ \hline
$a/R_{\s P}$ \textcolor{blue}{$^1$} & 16.51	  &  36.92  &	41.55	& 48.99	 & 55.00 \\

$a/R_{\s P}$ \textcolor{blue}{$^2$} & 16.50  & 35.70 & 40.98 & 48.61 & 54.47 \\
$a/R_{\s P}$ \textcolor{blue}{$^3$} & --     & 35.91 & 40.99 & 48.64 & 54.49 \\
$e\times 10^{-3}$ \textcolor{blue}{$^1$} & 0.58	  &  36.70  &	16.78	& 6.82	 & 12.22 \\
$e\times 10^{-3}$ \textcolor{blue}{$^2$} & 0.05 & 0.01 & 0.00 & 0.00 & 5.54 \\
$e\times 10^{-3}$ \textcolor{blue}{$^3$} & -- & 5.8  & 2.0  & 3.3 & 5.9 \\
inc $(^\circ)$ \textcolor{blue}{$^1$} & 112.90  &  112.86 &	112.92	& 112.57 & 113.09 \\
inc $(^\circ)$ \textcolor{blue}{$^2$} & 0.0  & 0.0 & 0.0 & 0.4 & 0.3 \\
inc $(^\circ)$ \textcolor{blue}{$^3$} & --   & 0.81 & 0.13 & 0.39 & 0.24 \\
\hline
$M$ $(^\circ)$      \textcolor{blue}{$^1$} & 356.76 &  18.79  &  	49.94 &	326.33  &    32.45 \\
$\omega (^\circ)$ \textcolor{blue}{$^1$} & 293.25 &  348.42 & 	293.25&	293.25  &    293.25 \\
$\Omega (^\circ)$ \textcolor{blue}{$^1$} & 227.40 &  227.42 & 	227.45&	227.09  &    227.16 \\
\hline
Mass (kg) \textcolor{blue}{$^2$}& $1.59 \times 10^{21}$ &  $1.49 \times 10^{15}$    &  $4.49 \times 10^{16}$ & $1.65 \times 10^{16}$ &  $4.79 \times 10^{16}$ \\
\hline
 Errors (km) \textcolor{blue}{$^2$}& 5  &  514	& 23 & 	135	& 29 \\
 $\Delta a$ (km)       \textcolor{blue}{$^1$}& 0.043	  &  1761	& 1086 & 	637	& 443\\
 $\Delta e$            \textcolor{blue}{$^1$}& 7.13$\times 10^{-5}$&	0.034	& 0.025	  &  0.018	& 0.016  \\
 $\Delta i$ $(^\circ)$    \textcolor{blue}{$^1$}& 7.26$\times 10^{-4}$&	0.08	& 0.12	  &  0.88	& 0.58  \\ 
 $\Delta a$ (km)         \textcolor{blue}{$^4$}& 1.$\times 10^{-3}$&	1792	& 1147 & 616	& 437\\
 $\Delta e$              \textcolor{blue}{$^4$}& 1.8$\times 10^{-7}$&	0.041	& 0.031	  & 0.015	& 0.015  \\
 $\Delta i$ $(^\circ)$      \textcolor{blue}{$^4$}& $\sim$0.00 &	0.08	& 0.12	  & 0.88	& 0.58  \\
\hline

\end{tabular}}\\
\textcolor{blue}{$^1$} From the JPL Horizons site, at epoch 2021/01/01.  (\url{https://ssd.jpl.nasa.gov/?horizons}) {choosing the ecliptic as reference plane.}\\
\textcolor{blue}{$^2$} From \cite{Brozovic+2015} {averaged mean orbital elements derived based on 200 years of orbital integration. The epoch for the elements is JED 2451544.5}. \textcolor{blue}{$^3$} From \cite{Showalter+2015} {, fitting a precessing ellipse. The epoch is Universal Coordinate Time (UTC) on 1 July 2011.} {$M,\omega,\Omega$ are mean anomaly, argument of pericenter, and longitude of the node respectively}. Mass of Styx has solution with null value but we adopted $1\sigma$ value. {Errors correspond to in-orbit uncertainties given in \citet{Brozovic+2015}. Orbital element variations ($\Delta a, \Delta e,$ and $\Delta i$) correspond to those extracted from JPL ephemerides\textcolor{blue}{$^1$} and to N-body integrations considering three body problem\textcolor{blue}{$^4$}, in 40 yr timespan.}
\end{center}
\vspace{-0.1 cm}
\end{table*}

The small moons describe their orbits with respect to the barycenter of the \PC binary in the nearly {circular and} coplanar orbital geometry. {Several works have studied the dynamical stability of the small moons and also considered the possibility of the existence of putative satellites} \citep[e.g.,][]{Weaver+2006, Kenyon+2019a, Kenyon+2019b}, some of them considering the moon's masses one order of magnitude larger than those determined by observational data (\citealt{Tholen+2008, Pires+2011, Youdin+2012}). The orbital periods of the small moons place them very close (depending on the uncertainties) even inside  the N/1 mean motion resonances (MMRs) with Charon, namely, 3/1, 4/1, 5/1, and 6/1, for Styx, Nix, Kerberos, and Hydra, respectively. {For example, \citet{Brozovic+2015} obtained the period ratios of a satellite and Charon, such as 3.1565, 3.8913, 5.0363, 5.9810, for Styx, Nix, Kerberos, and Hydra, respectively.} While the double synchronous state of the \PC binary is an indicator of the tidal evolution of the system, the positions of the moons with respect to the MMRs could be an indicator of smooth migration process. 

It is well accepted that Charon was formed as a result of a giant collision that most likely happened when the population of the Kuiper Belt was much denser than today (\citealt{Canup2005, Ward+2006, asphaug2006, Canup+2011, McKinnon+2017, Walsh+2015}). This giant impact could also originate the very small circumbinary satellites Styx, Nix, Kerberos, and Hydra. There is evidence that the satellites are of the age of Pluto; {indeed, the crater counting data from \textit{New Horizons} imply that the surface ages of Nix and Hydra are, at least, of 4 billion years \citep{Weaver+2016}}.

Formation theories for the small moons in the \PC system include an \textit{intact capture scenario} and a \textit{planetesimal capture scenario}. In the {scenario of intact capture} \citep[][]{Canup2005, Canup+2011}, the proto-Charon grows rapidly, in about 30\,h after collision event, within a massive debris swarm produced during the collision and extending from $4 R_{\s P}$ to $25 R_{\s P}$ (in units of the Pluto's radius). {The commonly assumed initial position of Charon in the disc is around $4 R_{\s P}$ \citep[e.g.,][]{Chen+2014a,Chen+2014b,Woo+2018}}. Depending on the characteristics of the impact, the initial orbit of Charon varies its form, from circular to highly eccentric one ($e_{\s C} \sim 0.50$).  The small moons belong to the debris swarm, and their initial positions are generally considered to be closer to Pluto than today.

To place the moons at their current locations, \citet{Ward+2006} proposed the mechanism of resonant transport. The main idea of the method is that, during the tidal expansion of the \PC orbit, the small satellites can be captured into \textit{resonances} with Charon and migrate outward together with Charon. Several authors have tested this hypothesis {and analysed the probability of capture in the MMRs} of the kind N/1 \citep[e.g.,][]{Lithwick+Wu2008, Chen+2014b, Woo+2018, Kenyon+2021}, which are considered to be most strong around binaries \citep[e.g.,][]{Cuello&Giuppone2019, Gallardo+2021}. However, \citet{Lithwick+2008} have found some difficulties to adjust the values of the Charon’s eccentricity, $e_{\s C}$, such that,  in order to transport safely Nix to the 4/1 MMR,  it should be $e_{\s C}<0.024$, while, in order to transport Hydra to the 6/1 MMR, it should be $e_{\s C}>0.04$.  {\citet{Chen+2014b} have found stable solutions (that is, not ejected from the system) for the test particles at the $5/1$, $6/1$, and $7/1$ MMRs, but none at the $3/1$ and $4/1$ MMRs. Moreover, the orbits of the  surviving particles were highly eccentric, in contrast with the currently nearly circular orbits of the small moons. In addition, the authors have found that, when the hydrostatic value of Pluto’s zonal harmonic, $J_2$, was included in the model, there was no stable transport at the regions near the N/1 MMRs.}

To overcome the problems of the resonant transport model, the scenario of planetesimal capture gained more attention. This scenario considers a ring of ejected material, which is ranging up to $60 R_P$, or even $200 R_P$, more than that in the intact capture scenario \citep[e.g.,][]{Pires+2012,Desch+2015, Walsh+2015,Kenyon+BromleyIII}. {\citet{Smullen+2017} studied the evolution of a debris disc resulting from the Charon-forming impact, established regions of stability according to tidal evolution of \PC binary, and characterised the collisions onto Charon’s surface that might leave visible craters.} \citet{Woo+2018} studied several possibilities for {survival of the test particles in the regions of the known moons} during the tidal expansion of the \PC binary (\textit{in-situ} formation scenario). Applying different tidal models, the authors found out most promising results when considered a constant-$\Delta t$ tidal model, with large dissipation coefficient ($A \sim 40$) and the initially circular orbit of Charon ($e_C=0.0$); however, in their simulations, they did not consider the impact of the zonal harmonic of the Pluto's oblateness.

 \begin{figure*}
  \centering
\mbox{\includegraphics[width=1\columnwidth]{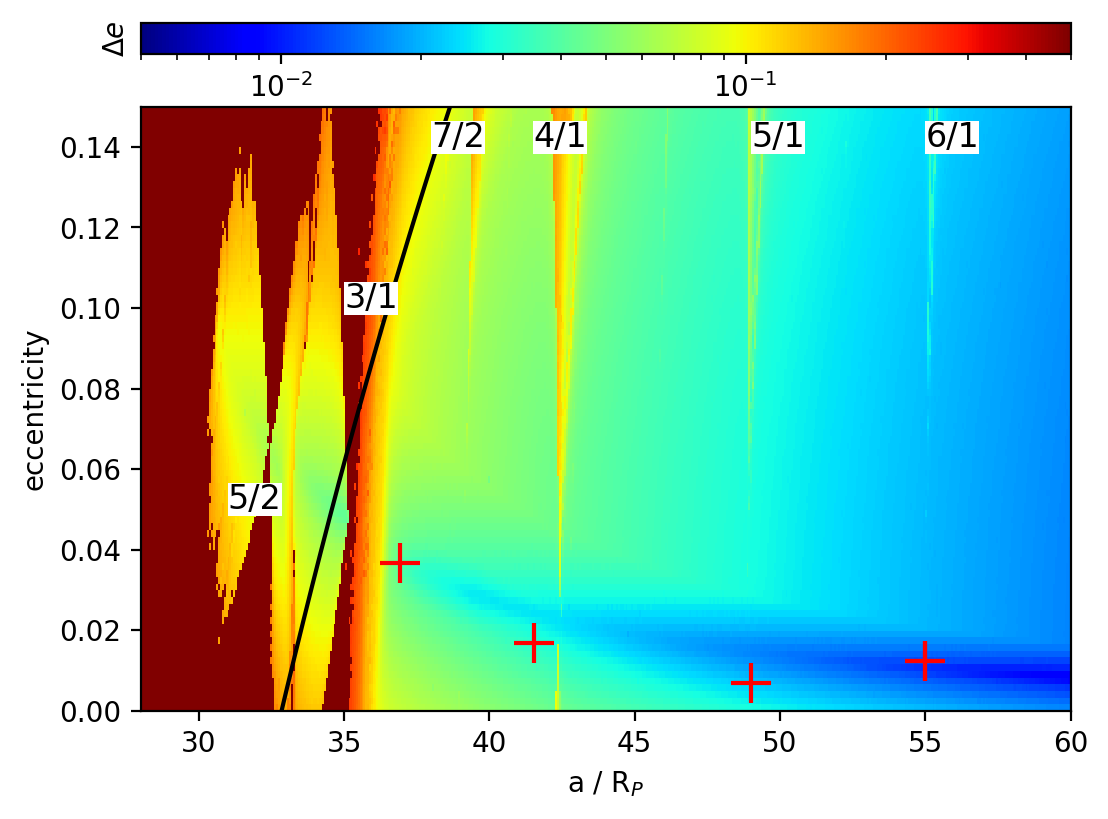}} \mbox{\includegraphics[width=1\columnwidth]{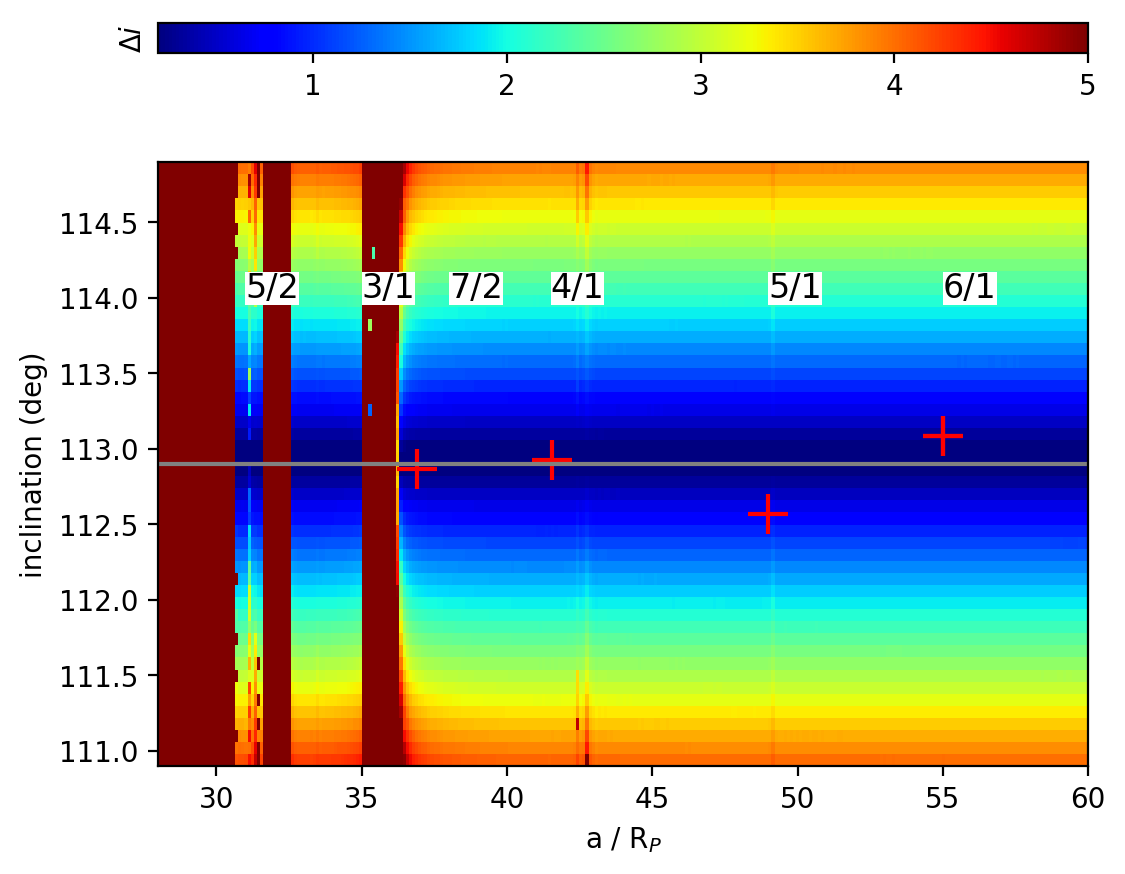}} \\
\caption{Dynamical maps on the $a$--$e$ plane (left panel) and $a$--$I$ plane (right panel) of the moon's initial conditions calculated for the \PC binary and a small moon of the Styx's mass {with initial Styx's orbital elements from JPL in Table \ref{tab.Pluto}}. The current positions of the moons are shown by red crosses. The locations of the main MMRs are indicated, {and the gray line at right panel indicate the current inclination of Charon}.}
  \label{fig1-conservative}
  \vspace{-1em}
  \end{figure*}

The results obtained either by the tidal transport or \textit{in situ} formation scenarios are still inconclusive due to the poor knowledge of the precise orbital elements and masses of the small moons. The most precise orbital parameters {of the \PC system}, reported in \cite{Brozovic+2015} and \cite{Showalter+2015} (see Table \ref{tab.Pluto}), present {dispersion} of several orders of magnitude in the eccentricities and inclinations of the small moons. {The solution from \citet{Showalter+2015} locked Pluto and Charon to the ephemerides of JPL, PLU043}. Our data are taken from the \textit{JPL-Horizons} and correspond to a pre-computed solution PLU058/DE440, a fit to ground-based, HST, and New Horizons spacecraft encounter astrometry in the interval 1965-2018. To illustrate the uncertainties in orbital fits we show in Table \ref{tab.Pluto} the in-orbit errors (along the track) reported by \citet{Brozovic+2015} and also include the amplitude observed in the variation of orbital elements (see also the Figure \ref{fig-evol}). The precise orbital dynamics of the small moons depends on the initial osculating orbital elements {and the masses} of the moons. 


With this in mind, we introduce an additional mechanism, which could provide a robust explanation for the existence of the four small circumbinary satellites at their current positions. For this, we first give a global view of the dynamics of the current Pluto's system  of small satellites in Sect.\,\ref{sec.conservative}. In Sect.\,\ref{sec.tidalmodel}, we present the model, which describes the tidal interactions between Pluto and Charon, and discuss the choice of the parameter values adopted in this paper. In the next, we analyse how the tidal expansion of the \PC binary, starting at the different initial configurations, could affect the behaviour o the small moons located at their current positions (Sect.\,\ref{PC+moon}). In Sect.\,\ref{sec.moons},  we analyse, in the frame of the {\it in-situ} formation scenario, the effects of tidal evolution of the Charon's orbit  on a large grid of the parameter values and the initial conditions. In that section, we also study the impact of the zonal harmonic of the Pluto's oblateness on the moon's dynamics. To overcome the problem of survival of the moons during the passages through the low-order MMRs with Charon,  we investigate the effects of a mass accreting Charon on the moon's behaviour in Sect.\,\ref{sec.massgrowth}. Finally, we present our conclusions in Sect.\,\ref{sec.conclusions}.


\section{Dynamical portrait of the current \PC system}
\label{sec.conservative}

It is generally accepted that the current orbital configuration of the \PC binary and the four small moons is a product of the evolutionary history of the whole system during its lifetime, since the collision event that gave rise to the Pluto's satellites. In this context, the detailed analysis of the current relative positions of the Pluto's system members may provide important constraints on the dynamical past of the whole system.  

{A representative} picture of the \PC system is shown in Fig.\,\ref{fig1-conservative}; it is {calculated with the data from the \it{JPL Horizons} site (see Table \ref{tab.Pluto})}. To construct it,  we use a modified Jacobi reference frame, in which orbital elements of {the} small moon are referred to the centre of mass of the PC system, while the mutual interactions between the moons are neglected\footnote{{\citet{Pires+2011} have reported gravitational effects, due to Nix and Hydra, on the test particles in the external region of the \PC binary, but the masses of the moons used were one order larger than those reduced from the observations.}}.  In this reference frame, the motion of a particle of the Styx's mass is simulated through the numerical integration over $200\,yr \; (\sim$ 3500 Styx's orbital periods), and subsequently analysed using $\Delta e$ or $\Delta i$ indicators, {to better address the structure of the resonances. We calculate $\Delta e$ as the amplitude of maximum variation of the orbital eccentricity of the satellites during the integrations, $\Delta e= \left(e_{\mathrm{max}}-e_{\mathrm{min}}\right)$. Similarly,  $\Delta i= \left(i_{\mathrm{max}}-i_{\mathrm{min}}\right)$ and $\Delta a= \left(a_{\mathrm{max}}-a_{\mathrm{min}}\right)$}. 
Regions with higher $\Delta e$ lead to chaotic motion and the indicator also serves to identify separatrices of mean motion resonances, being a powerful indicator of secular and resonant dynamics
\citep[e.g.,][]{Dvorak.etal.2004,Ramos.etal.2015}.

The instabilities in the motion of the particles are represented using colour scale on the maps in Fig.\,\ref{fig1-conservative}, on the $a$--$e$ plane (left panel) and $a$--$I$ plane (right panel) of the initial conditions, where $a$, $e$ and $I$ are the semimajor axis, the eccentricity and the inclination of the particle, respectively.  In the stability maps the initial conditions for Charon are given in Table \ref{tab.Pluto}, {whereas the initial values of the angular variables of the circumbinary moon are those given to Styx in Table \ref{tab.Pluto}}. The colour scale varies with {$\Delta e$ and $\Delta i$}, increasing from blue (regular motion) to red (chaotic motion), as shown in the colour bar on the top of the figure. The positions of the four moons, Styx, Nix, Kerberos, and Hydra, are shown by the red crosses on the maps (see Table \ref{tab-init}). We note that all moons lie beyond the black continuous curve on the left panel, which delimits the domain of stable motion, according to the Holman--Wiegert criterion (HWC) derived in \cite{Holman+Wiegert1999} for circumbinary configurations, {and also noticed by \citet{Smullen+2017}}. 

Figure \ref{fig1-conservative} shows that the orbits of the moons are nearly circular, nearly planar and nearly resonant. Although these features of the Pluto's system have been already pointed out in the previous studies, it is still worth analysing them  in more detail. Indeed, as seen on the stability maps on the $a$--$e$ and $a$--$I$ planes, the small satellites are confined to the very small eccentricity and inclination regions of regular motion (blue colour). According to secular perturbation theories, very low eccentricities and inclinations are characteristic of the dynamical systems with Angular Momentum Deficit (AMD) close to zero \citep{michtchenko+rodriguez2011}. Regarding  initially excited dynamical systems (in this case, due to the collisional origin of Charon), the low-AMD configurations are generally attained by the systems evolving under weak and slow  dissipation. 

The interpretation of the near resonant distribution of the small satellites requires the analysis of an additional dynamical mechanism that is a capture into MMRs. The stability maps in Fig.\,\ref{fig1-conservative}  show the MMRs produced by Charon as vertical strips of chaotic motion (red colour). The low-order strong MMRs, such as 3/2, 5/3, 2/1, 5/2 and 3/1, are located inside the domain of instability defined by HWC (black curve on the left panel). In the low-eccentricity region, the 3/1 MMR can be regarded as an inferior limit of stability of the motion, and  Styx is stuck to it.  This moon lies so close to the 3/1 MMR that uncertainties in determination of the angular variables of its orbit, such as the mean longitude and the longitude of the pericenter, might put it {closer to} the chaotic layer of the 3/1 MMR {(e.g., in-orbit and radial errors reported in \citealt{Brozovic+2015})}. The same is true for Nix, with respect to the 4/1 MMR {(see panels Styx and Nix in Figure \ref{fig-mapaM})}, but {most noticeable} for Kerberos and Hydra, with respect to the 5/1 and 6/1 MMRs, respectively {(see bottom panels in Figure \ref{fig-mapaM})}. This fact suggests that the small moons could be involved in one of the MMRs at some period of their common past history. Therefore, in this paper, we will analyse the interaction of the small moons with the main MMRs during the tidal expansion of the Charon's orbit.  

It is worth emphasising that the uncertainties in the determination of the moon's masses and the actual positions can provide a general view of the phase space of the \PC system, but not specifying whether the objects are evolving nearby or inside the MMRs.

\section{Tidal model}\label{sec.tidalmodel} 

In this section, we describe the tidal model, which we apply to investigate the orbital expansion of Charon and the implications for the orbital motion of the small moons. As described in the previous section, the orbital configuration of the Pluto's satellites is better represented in terms of the modified Jacobi elements. The initial Jacobi orbital elements are transformed to the Jacobi rectangular coordinates and velocities in a three-dimensional reference frame centred in Pluto. The equations of motion are solved using a N-body integrator with adaptable step-size (for details, see \citet{Rodriguez+2013}). We consider the system consisting of Pluto, Charon and an external small moon, where two large bodies undergo mutual tidal interaction, while the external moon is not (directly) tidally affected,  due to its small size and mass. We adopt a classical linear tidal model, in which the deformations of the bodies are delayed by a constant $\Delta t$, usually referred to as tidal time lag \citep{1979M&P....20..301M}. In this scenario, the \PC tidal interaction is modelled following \citet{Chen+2014a}\footnote{We refer the reader to this work for a complete analysis of the \PC tidal evolution.}, which assumes the collisional origin of Charon, when Charon emerges with the non-zero orbital eccentricity, $e_C$, and the initial semimajor axis $a_C= 4 R_{\s P}$ \citep{Ward+2006}.


Both Pluto and Charon would be spinning after the collision, with the spin axes oriented in some directions with respect to the orbital plane of the binary.
Due to the relatively small size and mass of Charon, in comparison with Pluto, the contribution of its spin to the total angular momentum is small, and it evolves quickly (less than 10 years) to the asymptotic tidal spin rate. For sake of simplicity, in this work, we assume zero obliquity for both Pluto and Charon; the model of the \PC binary, accounting for the  arbitrary obliquities and the Sun's perturbations, is studied in \citet{Correia+2020}.

The total angular moment of the \PC binary can be written, in the chosen reference frame and with the assumed approximations, as 
\begin{equation}
L = \mathcal{C}_{\s P} \Omega_{\s P} + \mathcal{C}_{\s C} \Omega_{\s C} + M_{PC} n_{\s C} a^2_{\s C} \sqrt{1-e_{\s C}^2},
\label{eq.angmot}
\end{equation}
where $M_{PC}=m_{\s P}m_{\s C}/(m_{\s P}+m_{\s C})$ and $n_C$ are the reduced mass of the binary and the mean motion of Charon, respectively. According to classical theories, $L$ is conserved during the tidal expansion of the \PC binary. Hence, its amount for the \PC system can be obtained knowing the masses of Pluto and Charon, $m_{\s P}$ and $m_{\s C}$, and the current values $a_{\s C}$ and $n_{\s C}$. Indeed, at dual synchronous state of the current \PC binary, the Charon's orbit is circularised, that is, we can assume  $e_{\s C}=0$, and  $\Omega_{\s P}$=$\Omega_{\s C}$=$n_{\s C}$. 

\begin{figure*}
\includegraphics[width=0.95\textwidth,angle=0]{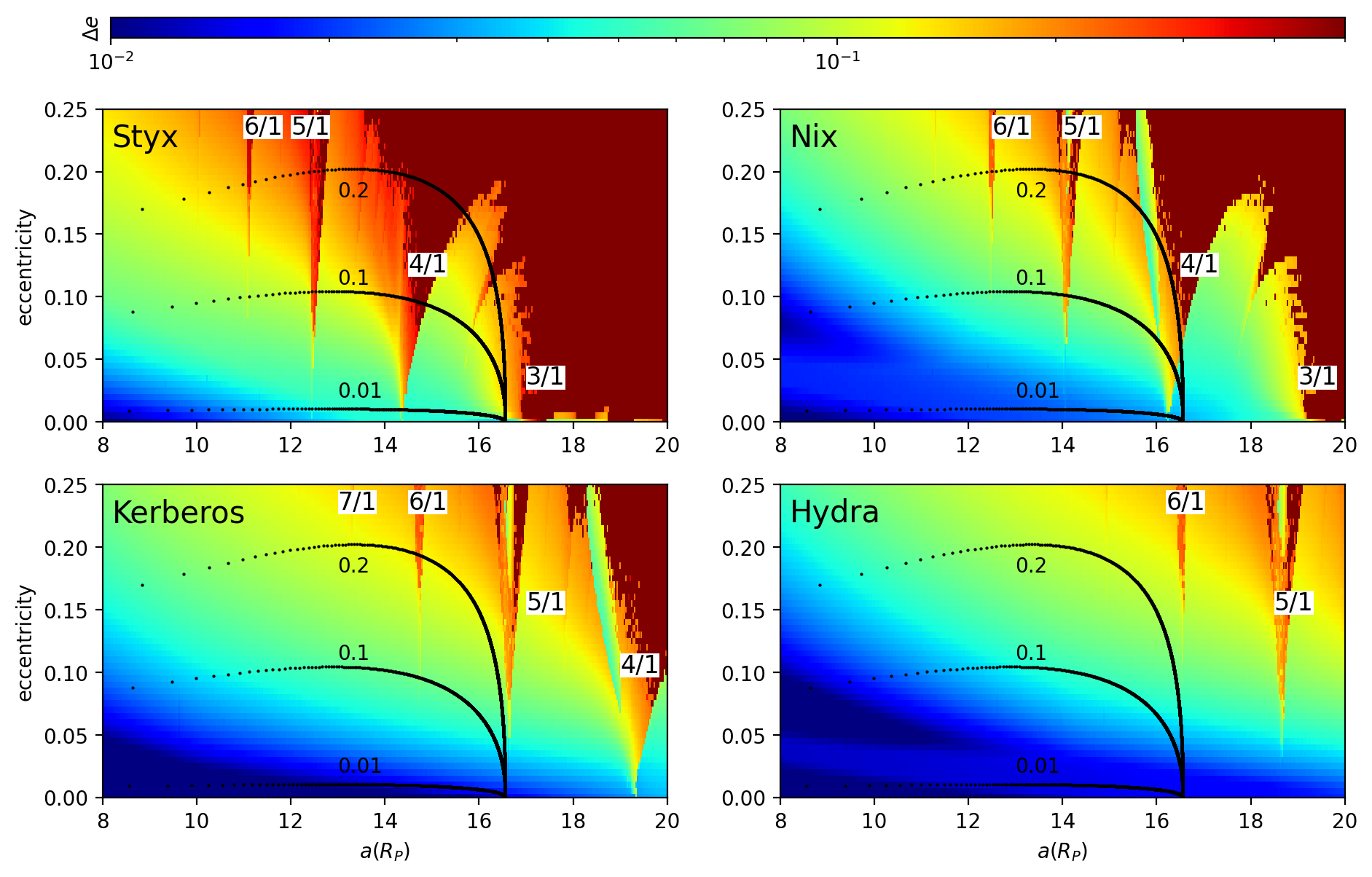}
\caption{Dynamical maps of Styx, Nix, Kerberos and Hydra on the $a_{\s C}$--$e_{\s C}$ plane of the Charon's initial conditions. On each map, black curves are the Charon's tidal trajectories, obtained for the initial $e_{\s C}$-values of 0.01, 0.1 and 0.2; the initial conditions of the corresponding moon are fixed {at those given in JPL from Table \ref{tab.Pluto}}. The blue tones represent regular motions, while the {brick} tones correspond to increasing instabilities and chaotic motion; the colour bar on the top shows {$\Delta e$}. The locations of the main MMRs are indicated.
}
\label{mapa-a1-e1}
\end{figure*}

Then, by substitution of these conditions into Eq.\,(\ref{eq.angmot}), the $L$ value is obtained; in the following, this value is used to obtain the angular velocity of Pluto, $\Omega_{\s P}$, resolving Eq.\,(\ref{eq.angmot}) for given starting values of $a_{\s C}$, $e_{\s C}$, and $\Omega_{\s C}$, the angular velocity of Charon's rotation.

{Following \citet{Chen+2014a}, we define the {dissipation} parameter $A$ given by}
\begin{equation}\label{eqA}
A=\frac{k_{\s 2C}}{k_{\s 2P}}\frac{\Delta t_{\s C}}{\Delta t_{\s P}}\Big{(}\frac{m_{\s P}}{m_{\s C}}\Big{)}^2\Big{(}\frac{R_{\s C}}{R_{\s P}}\Big{)}^5,
\end{equation}
where ($k_{\s 2C},k_{\s 2P}$) are the second order Love numbers and ($R_{\s P}$,$R_{\s C}$) are the radii for Pluto and Charon, respectively. We note that, for the masses and the radii of Pluto and Charon {($R_{\s C}=606$ km)},
\begin{equation}\label{dissipation}
\Bigg{(}\frac{m_{\s P}}{m_{\s C}}\Bigg{)}^2\Bigg{(}\frac{R_{\s C}}{R_{\s P}}\Bigg{)}^5\simeq2; 
\end{equation} 
thus, the constant $A$ can be closely understood as the ratio of time lags between Charon and Pluto. \cite{Chen+2014a} also have shown that, for an appropriate value of the ratio ($A\sim10$),  the eccentricity  of Charon's orbit, $e_{\s C}$, remains roughly constant during most of the tidal evolution.

Our current knowledge on the physical properties of the PC system does not allows us to constrain the range of $A$ values (see Eq. (\ref{eqA}) and \citet{Lainey2016,Quillen+2017}). However, it is easy to show through classical tidal theories that, for $A<<1$, tidal forces on Pluto dominate the tidal evolution of the binary, whereas, for $A>>1$, the orbital expansion and circularization are dictated by the tides on Charon. Previous works adopted different ranges for the possible values of $A$, namely,  $1<A<18$ \citep[][]{Chen+2014a,Chen+2014b}; $A=10$ or $A=40$ \citep[][]{Woo+2018}; $1<A<16$ \citep[][]{Correia+2020}. In this work, we initially explore the wide range of 0.01 < A < 100; however, in order to place test moons at the current Styx position, we adopt the range of $8<A<40$.

\citet{Chen+2014a} have shown the tidal evolution of \PC for different initial eccentricities of Charon. Here, we summarise the common features observed in the numerical integrations, which we will refer in our discussions below: (i) independently of the initial values of the Charon's eccentricity, $e_{\s C}$, {it presents a peak at some moment during the \PC expansion}, (ii) the time evolution of $\Omega_{\s P}/n$  is correlated with the peak of $e_{\s C}$ and can reach values of $\sim 10$, (iii) the semimajor axis of Charon attains its current value ($16.5R_{\s P}$) and the orbit is circularised, whereas both rotations synchronise with the orbital period {after around $3\times10^6$ yr}. It is worth emphasising that, {according the tidal evolution theories}, the double synchronous rotation is only attained when the orbit is completely circularised \citep[see][]{Chen+2014a}. Also, at the end of the evolution, the angular momentum exchange {between Pluto and Charon} ceases and both orbital and rotational components attain their equilibrium configurations in the system.

\section{Dynamics of the moons in the tidally evolving \PC system}\label{PC+moon}

In this section, we investigate the dynamics of the small moons located at their current positions, during the tidal expansion of the Charon's orbit. For this, we use the same techniques described in Sec.\,\ref{sec.conservative}. Considering that the Charon's orbital elements vary during the migration, we map the neighbourhood of each moon applying the conservative model, for different values of the semimajor axis, $a_{\s C}$, and the eccentricity, $e_{\s C}$, of Charon's orbit and the fixed mass and initial conditions of the satellite (Table\,\ref{tab.Pluto}). In this way, we can visualise the dynamical behaviour of each moon at different configurations of the \PC binary acquired  in the past. 
Figure \ref{mapa-a1-e1} shows the dynamical maps on the $a_{\s C}$--$e_{\s C}$ planes, obtained for {current locations of} Styx, Nix, Kerberos and Hydra {retrieved from JPL-Horizons}. The black curves show the positions of Charon on these planes, as they were obtained through simulations of its tidal expansion in the past (described in the previous section). Three values of the Charon eccentricity were used, namely 0.01, 0.1 and 0.2, while the other parameters were fixed at $a_{\s C}/R_{\s P}=4$, $\Omega_{\s C}/n_{\s C}=2$, $A=10$ and $\Delta t_{\s P}=600s$. By construction, the three paths converge to the current position of Charon, at $a_{\s C}\cong 16.5R_{\s P}$ and $e_{\s C}\cong 0$.  

Now, we can investigate the dynamical behaviour of each moon according to the position of Charon,  at any moment of its past history. Indeed, for a chosen position of Charon in one of its paths on the $a_{\s C}$--$e_{\s C}$ planes  in Fig.\,\ref{mapa-a1-e1}, we can identify dynamical features, which affect the behaviour of the moon at that instant.  These features are mainly MMRs, which are identified by plotting the moon's dynamics with the colour scale. In this way, we can distinguish between the stable (blue-white) and unstable (red) behaviour of satellites. The instabilities of the motion are clearly related to the MMRs, whose locations are indicated by the corresponding ratio in Fig.\,\ref{mapa-a1-e1}.  

The most strong MMRs are of low order, 3/1 and 4/1, and we expect that the dynamical stability of the moons would be significantly affected by these resonances. Figure \ref{mapa-a1-e1} shows that Charon's paths (black curves) never cross  the 3/1 MMR and, consequently, none of the four small moons suffer the destabilising effects of this resonance. Currently, Styx  stays close to this resonances (left-top panel), but its motion is still stable. It is worth noting here that the 3/1 MMR can be considered as an inferior limit of stability of the circumbinary motion, as shown in Fig.\,\ref{fig1-conservative}.

According to Fig.\,\ref{mapa-a1-e1}, the actual position of Styx has been crossed by the 4/1 MMR at some moment in the past, also as of Nix (right-top panel) at another period. Our simulations of the moon's dynamics (described in the next section) show that the perturbations produced by this resonance increase with the increasing eccentricity of Charon.  For Charon's paths with $e_{\s C}=0.1$ and $0.2$, Styx and Nix spend a time evolving in the 4/1 MMR, which might be sufficient to strongly excite their eccentricities and even eject from the current positions. On the contrary, for Charon on quasi-circular orbit ($e_{\s C}=0.01$), the passages of Styx and Nix through the 4/1 MMR may occur quickly and without significant excitation of their motions. 

The width and, consequently, the dynamical effects of the higher order MMRs, such as 5/1, 6/1, 7/1, etc, rapidly decreases with the decreasing order, as seen in  Fig.\,\ref{mapa-a1-e1}.  We expect that their passages through current positions of the small moons would not produce perturbations to destabilise their motion, particularly, for Charon evolving on the quasi-circular orbit during these passages.  Thus, we conclude that, in order to ensure the simultaneous orbital stability of the four small moons during the orbital expansion of the \PC binary, Charon's eccentricity should be constrained to lower values. This fact points towards a dynamical scenario, in which the adequate choice of Charon's eccentricity at the beginning of the tidal expansion is essential, in order to guarantee the orbital stability of the system of moons. In the next section, we will simulate numerically the impact of the migrating Charon on the orbital motions of Styx, Nix, Kerberos and Hydra by taking into account simultaneous effects of the tidal evolution and gravitational perturbations.

\section{Simulations of the moon's behaviour under the tidal expansion of the Charon's orbit}\label{sec.moons}

In this section, we analyse the behaviour of Styx, Nix, Kerberos and Hydra during the orbital expansion of Charon's orbit and their passages through the MMRs. Here, we assume the \textit{in-situ} formation of the small moons, that is, our numerical simulations start with the satellites at their current positions. We work in the modified Jacobi reference frame, where small moons are orbiting the barycenter of the \PC binary. {In this frame, we neglect the mutual perturbation between the moons, due to their small masses, that allows us to solve their orbital motions in a separate way. We solve the exact equations of motion for each satellite, according to the tidal model described in Sect.\,\ref{sec.tidalmodel} and following \citet{Rodriguez+2013}}. {We consider coplanar configurations and the initial mean anomalies, and longitudes of pericenter are set to zero.}

\begin{table}
\begin{center}
\caption{Parameters adopted in the numerical simulations. Here, $n_{\s C}$ is the mean motion of Charon. We note that the initial value of $\Omega_{\s P}$ is calculated taking into account the conservation of the total angular momentum of the system (see Eq.\,\ref{eq.angmot} ). The adopted parameters covers a broad range of initial conditions, as also suggested in the previous works \citep[e.g.,][]{Canup2005,Canup+2011,Chen+2014a,Chen+2014b,Correia+2020}}
\label{tab.PC}
\begin{tabular}{c|c}
\hline
Parameter & Values \\ \hline
$a_{\s C}/R_{\s P}$ & 3, 4, 5, 6, 7\\
$e_{\s C}$ & 0, 0.001, 0.01, 0.1, 0.2\\
$\Omega_{\s C}/n_{\s C}$ & 0.5, 1.5, 2, 3, 4\\
$A$ & 0.01, 0.1, 1, 10, 100\\
$\Delta t_{\s P}$ (s) & 6, 60, 600\\
\hline
\end{tabular}
\end{center}
\end{table}

We perform a set of numerical simulations varying the main parameters of the problem, namely, the initial semimajor axis and eccentricity of Charon's orbit {($a_{\s C},e_{\s C}$)}, the initial rotation velocities of Pluto and Charon ($\Omega_{\s C}$ and $\Omega_{\s P}$), the time lag of Pluto $\Delta t_{\s P}$, and the constant of dissipation $A$ (see Table \ref{tab.PC}). {The initial orbital angles of Charon are set to zero.} We fix a fiducial set of parameters, such that $a_{\s C}/R_{\s P}=4$, $e_{\s C}=0.001$, $\Omega_{\s C}/n_{\s C}=2$, $A=10$ and $\Delta t_{\s P}=600s$. Then, the simulations are done varying the parameters one-by-one, choosing the values listed in Table \ref{tab.PC}. 
{It is worth noting that the initial value of $\Omega_{\s P}$ is not a free parameter, but is determined by the conservation of the angular momentum of the binary through  Eq.\,(\ref{eq.angmot}).}

The masses and current semimajor axes of the external moons are taken from \cite{Brozovic+2015} (see Table \ref{tab.Pluto}). The initial orbital eccentricities {of the moons} are assumed to be almost circular ($e=10^{-5}$).  Note that, in this section, we work in the frame of the planar problem, when the small moon evolves on the orbital plane of the \PC binary.
Physical parameters, such as the masses and radii of Pluto and Charon, are the same described in Sect.\,\ref{sec.tidalmodel}. 
Since the tidal evolution of the \PC binary is not affected by the external small moon, we do not show the variations of their orbital and rotational components {in this section} {\citep[see Fig.\,2 in][]{Chen+2014a}}. In addition, we performed simulations for both $\Delta t_{\s P}=0.06$\,s and $\Delta t_{\s P}=0.6$\,s, but the results obtained were discarded, since Charon did not reach its current semimajor axis after 100 Myr in these cases. We integrate the equations of motions over\,100 Myr for $\Delta t_{\s P}=6$\,s and $\Delta t_{\s P}=60$\,s; for $\Delta t_{\s P}=600$\,s, we {consider} 10\,Myr {as the integration timescale}. 

In the following, we summarise the results obtained for each of the four moons, emphasising the cases in which the final orbital configuration of the small moon is similar to its current one. {We note that, despite the fact that the moons are currently in the nearly resonant configurations, the uncertainties in their masses and orbital parameters could still place them inside the MMRs (see our discussion in Sect. \ref{sec:intro} and the Fig. \ref{fig-mapaM}). In Table \ref{tab-results} we include the cases, for which each of the moons reaches the end of simulation in stable motion. 
Additionally, we identify the initial conditions, for which the final eccentricity of the moon is smaller than $0.05$ that is compatible with the current eccentricities of the moons (see our Fig. \ref{fig-evol}).}


\begin{figure}
\includegraphics[width=0.49\columnwidth]{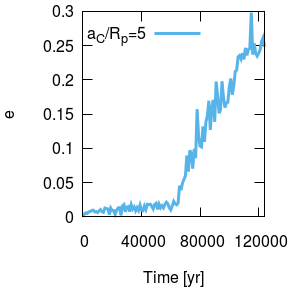}
\includegraphics[width=0.49\columnwidth]{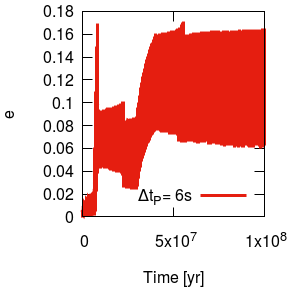}
\caption{ Two typical examples of the Styx orbital evolution resulting in either instability (left panel) or the final configuration, which is significantly different from its current orbit (right panel). \textit{Left panel}: The initial conditions used are $a_{\s C}/R_{\s P}=5$, $e_{\s C}=0.001$, $\Omega_{\s C}/n_{\s C}=2$, $A=10$ and $\Delta t_{\s P}=600$\,s. Styx is  ultimately ejected from the system when {the system crosses the 4/1 MMR} between Charon and Styx (not shown), resulting in a strong increase of the Styx's eccentricity. \textit{Right panel}: The initial conditions used are $a_{\s C}/R_{\s P}=4$, $e_{\s C}=0.001$, $\Omega_{\s C}/n_{\s C}=2$, $A=10$ and $\Delta t_{\s P}=6$\,s. The orbital eccentricity is excited, {due to the crossing of the 4/1 and 7/2 MMRs (not shown)}, to attain the mean value $\simeq0.1$, that is substantially larger than the current Styx's eccentricity.}
\label{styx-figures1}
\end{figure}

\subsection{Styx}

Styx is {located} at a mean distance of 42,650 km ($\simeq35.9 R_{\s P}$) from the barycentre of the \PC binary, and is close to the 3/1 MMR with Charon (see Fig.\,\ref{fig1-conservative} and Fig.\,\ref{fig-mapaM}).
{Other initial conditions different from those shown in Table \ref{tab-results}, tested for this satellite,} lead to either ejection of the small moon from the system (Fig.\,\ref{styx-figures1} left panel) or a final orbit, which is different from the current one (Fig.\,\ref{styx-figures1} right panel). In most cases, instabilities occur when Styx approaches the 4/1 or 5/1 MMRs, in this last case, only for high initial $e_{\s C}$. 



\begin{figure}
\includegraphics[width=0.49\columnwidth]{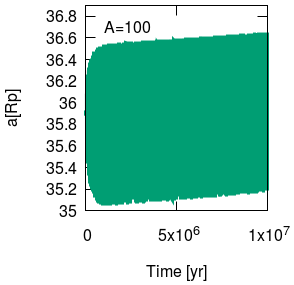}
\includegraphics[width=0.49\columnwidth]{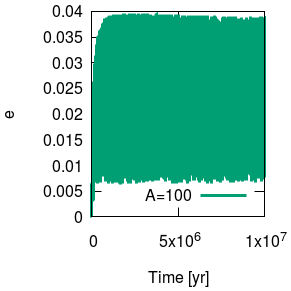}\\
\includegraphics[width=0.49\columnwidth]{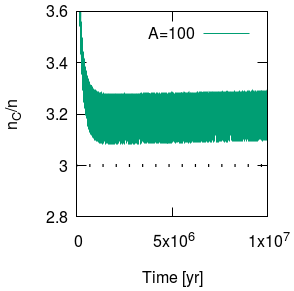}
\includegraphics[width=0.49\columnwidth]{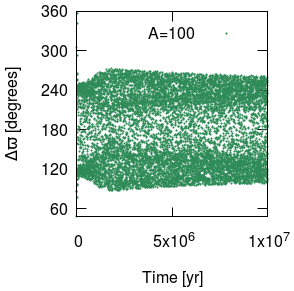}
\caption{\textit{Top row}: Time evolution of the Styx's semimajor axis (left) and eccentricity (right) during the tidal expansion of the \PC binary. The initial conditions used in the simulations are  $a_{\s C}/R_{\s P}=4$, $e_{s C}=0.001$, $\Omega_{\s C}/n_{\s C}=2$, $A=100$ and $\Delta t_{\s P}=600$ s. The final values of the orbital elements remain close to the current ones. 
\textit{Bottom row}: Time evolution of the mean-motion ratio (left) and $\Delta\varpi$ (right). Note that the mean value of $n_{\s C}/n$ is larger than 3. The secular angle $\Delta\varpi$ oscillates around $180^{\circ}$.
}
\label{styx-figures2}
\end{figure}

Figure \ref{styx-figures2} illustrates the Styx's orbital evolution obtained for $A=100$. We note that, despite an apparently slow increase, the semimajor axis of Styx remains close to the current value. The eccentricity acquires a mean value of 0.02, varying roughly between 0.007 and 0.04, which is also consistent with the oscillation of the current Styx's eccentricity. The final value of the mean motion ratio is larger than the nominal location of the 3/1 MMR (see Fig.\,\ref{fig1-conservative}); the difference between the pericenter longitudes of Styx and Charon, $\Delta\varpi$, oscillates around $180^{\circ}$, {which} is an indicator of the dissipative migration (see Sect.\,\ref{PC+moon}). We have checked that, for this particular simulation, all critical angles associated to the 3/1 MMR circulate. 

The results obtained for Styx point out that the scenario with the initially quasi-circular orbit of Charon and high dissipation ratio (large $A$) favours the stability of the Styx's motion during the orbital expansion of the \PC binary. However, we note in Table \ref{tab-results} that only {very specific combinations of the parameters allows} Styx to remain on the stable orbit {as} the 4/1 MMR with Charon is crossed. For most of the explored initial conditions, the Styx's motion is strongly excited by the passage through this resonance, resulting in the escape from the system. In Sect.\,\ref{sec.massgrowth}, we test the orbital stability of Styx (and the other small moons) assuming an on-time accretion of the Charon's mass.

\begin{table*}
\caption{{List of the parameters and initial conditions for which the four small moons survive in the numerical simulations. Values in bold correspond to the simulations resulting in a good agreement with the current orbital configuration of the small moons ($e<0.05$), according to the adopted criteria (see Sect.\,\ref{sec.moons})}}
\begin{center}
    \vspace*{0.3cm}
\begin{tabular}{c|c|c|c|c}
\hline
Parameter & Styx & Nix & Kerberos & Hydra \\ 
\hline
$a_{\s C}/R_{\s P}$ &  \textbf{4}  & 3, 4, 5, 6, 7 & \textbf{3}, \textbf{4}, \textbf{5}, \textbf{6}, \textbf{7} & \textbf{3}, \textbf{4}, \textbf{5}, \textbf{6}, \textbf{7}\\
\hline
$e_{\s C}$     & \textbf{0} & \textbf{0}, \textbf{0.001}, \textbf{0.01} & \textbf{0}, \textbf{0.001}, \textbf{0.01}, \textbf{0.1}, \textbf{0.2} & \textbf{0}, \textbf{0.001}, \textbf{0.01}, \textbf{0.1}, \textbf{0.2} \\
\hline
$\Omega_{\s C}/n_{\s C}$ &  1.5, 2 & 0.5, \textbf{1.5}, 2, \textbf{3}, \textbf{4} & \textbf{0.5}, \textbf{1.5}, \textbf{2}, \textbf{3}, \textbf{4} & \textbf{0.5}, \textbf{1.5}, \textbf{2}, \textbf{3}, \textbf{4} \\
\hline
$A$ & 10, \textbf{100} & \textbf{10}, 100 & \textbf{0.01}, \textbf{0.1}, \textbf{1}, \textbf{10}, \textbf{100} & \textbf{0.01}, \textbf{0.1}, \textbf{1}, \textbf{10}, \textbf{100} \\
\hline
$\Delta t_{\s P}$\,(s) & 6, \textbf{60}, 600 & \textbf{6}, 60, 600 & \textbf{6}, \textbf{60}, \textbf{600} & \textbf{6}, \textbf{60}, \textbf{600} \\
\hline 
\end{tabular}\label{tab-results}
\end{center}
\end{table*}

\begin{figure}
\includegraphics[width=0.49\columnwidth]{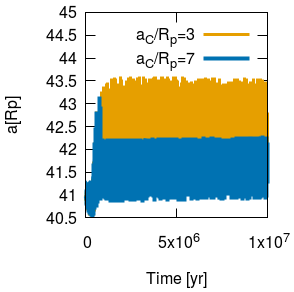}
\includegraphics[width=0.49\columnwidth]{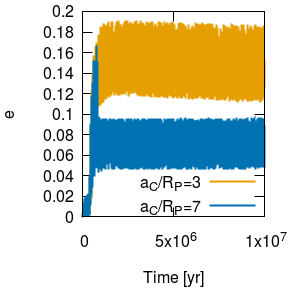}\\
\includegraphics[width=0.49\columnwidth]{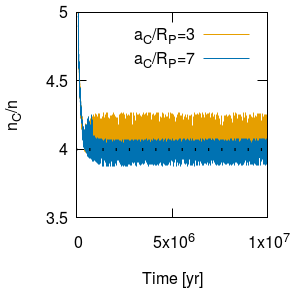}
\includegraphics[width=0.49\columnwidth]{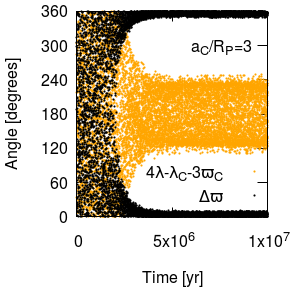}
\caption{\textit{Top row}: Time variations of the Nix's semimajor axis (left) and eccentricity (right), for two different initial values of $a_{\s C}$. The initial conditions are  $a_{\s C}/R_{\s P}=3$ and $a_{\s C}/R_{\s P}=7$, $e_{\s C}=0.001$, $\Omega_{\s C}/n_{\s C}=2$, $A=10$ and $\Delta t_{\s P}=600$\,s. For $a_{\s C}/R_{\s P}=3$, the eccentricity of Nix is excited to a mean value close to 0.16.
\textit{Bottom row}: Time evolution of the mean-motion ratio (left) and two angles for Nix (right), namely, the secular angle $\Delta\varpi$ and the 4/1 MMR critical angle $\phi_{4/1}=4\lambda-\lambda_{\s C}-3\varpi_{\s C}$. The mean value of $n_{\s C}/n$ is slightly smaller than 4, for $a_{\s C}/R_{\s P}=7$, whereas, for $a_{\s C}/R_{\s P}=3$, $\phi_{4/1}$ and $\Delta\varpi$ librate around $180^{\circ}$ and $0^{\circ}$, respectively.}
\label{nix-figures1}
\end{figure}

\subsection{Nix}

Nix lies at the mean distance of 48,690 km ($\simeq41R_{\s P}$) from the \PC barycenter and is currently close to the 4/1 MMR with Charon. The values of the initial conditions resulting in final configurations similar to the current orbit of Nix are shown in Table \ref{tab-results}. Figure \ref{nix-figures1}  shows the examples of the orbital evolution obtained for $a_{\s C}/R_{\s P}=3$ and $a_{\s C}/R_{\s P}=7$. For $a_{\s C}/R_{\s P}=3$ ({orange} curves), the resulting orbit clearly deviates from the current orbital configuration of Nix. In this case, the trapping into the 4/1 MMR excites the Nix's eccentricity, up to $e=0.16$. The final semimajor axis is larger than the current value, with a difference of $\simeq 2R_{\s P}$. {Moreover, the critical angle associated to the 4/1 MMR, $\phi_{4/1}=4\lambda-\lambda_{\s C}-3\varpi_{\s C}$, librates around $180^{\circ}$, as well all other angles associated to the same MMR}, while $\Delta\varpi$ oscillates around $0^{\circ}$ (Fig.\,\ref{nix-figures1}\,bottom-right).
{In this case, $e_C$ reaches $\simeq4\times10^{-5}$ at $10^7$ yr in one of simulations due to perturbations from Nix (not shown here).} 

On the other hand, for $a_{\s C}/R_{\s P}=7$ (blue curves),  the orbital evolution of Nix occurs with the eccentricity and semimajor axis close to their current values. All critical arguments associated to the 4/1 MMR circulate in {the case $a_{\s C}/R_{\s P}=7$}, {indicating} that the moon is out of the resonance, although is still very close to it. In this case, the mean motion ratio oscillates around a mean value slightly smaller than the nominal position of the 4/1 MMR (Fig.\,\ref{nix-figures1}\,bottom-left), in accordance with the current position of Nix given in \citet{Brozovic+2015} (see Figs.\,\ref{fig1-conservative} and \ref{mapa-a1-e1}).

The results shown in Table \ref{tab-results} indicate that, {in comparison with Styx, almost $40\%$ of the tested initial conditions result in the final orbital configurations, which are similar to the current orbit of Nix.}


\begin{figure}
\includegraphics[width=0.49\columnwidth]{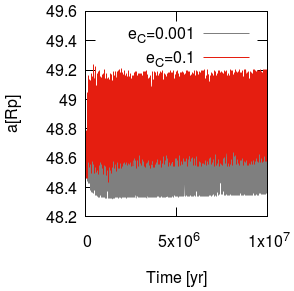}
\includegraphics[width=0.49\columnwidth]{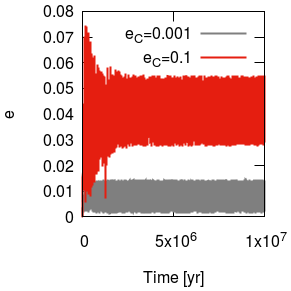}\\
\includegraphics[width=0.49\columnwidth]{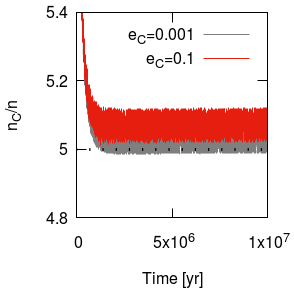}
\includegraphics[width=0.49\columnwidth]{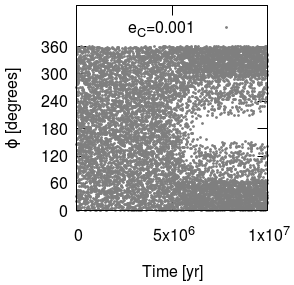}
\caption{\textit{Top row}: Orbital evolution of Kerberos for two initial conditions of $e_{\s C}$, namely, 0.001 and 0.1. The other parameters are $a_{\s C}/R=4$, $\Omega_{\s C}/n_{\s C}=2$, $A=10$ and $\Delta t_{\s P}=600$\,s. These simulations are successful, according to the adopted criteria (see Sect.\,\ref{sec.moons}.
\textit{Bottom row}: Time variations of the mean-motion ratio (left) and $\phi_{5/1}=5\lambda-\lambda_{\s C}-\varpi_{\s C}-3\varpi$ (right). We show the angle for the specific case of $e_{\s C}=0.001$. The mean value of $n_{\s C}/n$ is larger than 5.}
\label{kerb-figures1}
\end{figure}

\subsection{Kerberos}

Kerberos is the third small satellite orbiting the \PC binary, at a mean distance of $48.6 R_{\s P}$, that places it close to the 5/1 MMR with Charon. Table \ref{tab-results} shows {that all numerical simulations result in good agreement with the current orbit of this small moon.}



Figure \ref{kerb-figures1}  shows the orbital evolution of Kerberos during the tidal expansion of \PC binary for two values of the initial Charon's eccentricity, $e_{\s C}=0.001$ (grey curves) and $e_{\s C}=0.1$ (red curves). For both initial conditions, the output of the simulations fit satisfactory the current orbit  of the moon. All critical angular combinations of the 5/1 MMR circulate and the ratio $n_{\s C}/n$ oscillates around a mean value slightly larger than 5, that places the moon outside, but still close, the 5/1 MMR (see Fig.\,\ref{mapa-a1-e1}). 
We show the angle $\phi_{5/1}=5\lambda-\lambda_{\s C}-\varpi_{\s C}-3\varpi$, obtained for $e_{\s C}=0.001$, on the bottom-right panel in Fig.\,\ref{kerb-figures1}. At around $5\times10^6$ yr, this angle changes its behaviour from circulation to libration around $0^{\circ}$.

\subsection{Hydra}

\begin{figure}
\includegraphics[width=0.49\columnwidth]{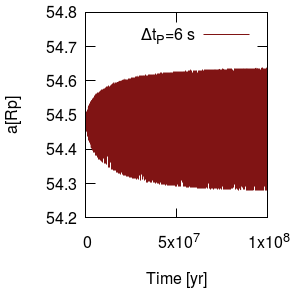}
\includegraphics[width=0.49\columnwidth]{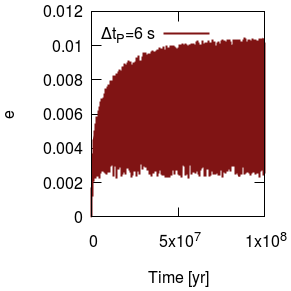}\\
\includegraphics[width=0.49\columnwidth]{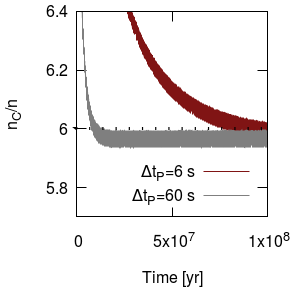}
\includegraphics[width=0.49\columnwidth]{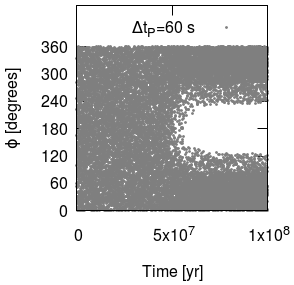}
\caption{\textit{Top row}: An example of the successful simulation concerning the orbital variation of semimajor axis (left) and eccentricity (right) of Hydra, according to the adopted criteria. The initial conditions are $a_{\s C}/R_{\s P}=4$, $e_{\s C}=0.001$, $\Omega_{\s C}/n_{\s C}=2$, $A=10$ and $\Delta t_{\s P}=6$\,s.
\textit{Bottom row}: Time variation of the mean-motion ratio (left) and $\phi_{6/1}=6\lambda-\lambda_{\s C}-\varpi_{\s C}-4\varpi$ (right) for two values of $\Delta t_{\s P}$, namely, 6\,s and 60\,s. We note that, for $\Delta t_{\s P}=60$\,s, $\phi_{6/1}$ {librates} with large amplitude around $0^{\circ}$. For $\Delta t_{\s P}=6$\,s, all critical angles associated to the 6/1 MMR circulate.}
\label{hydra-figures1}
\end{figure}

Hydra is the most distant known satellite of the \PC binary, with a current barycentric semimajor axis of $a \sim 54.5 R_P$. This small satellite is close to the 6/1 MMR with Charon. Most of the simulations reproduce the current orbit of Hydra, as shown in Table \ref{tab-results}. Figure \ref{hydra-figures1} shows the evolution for $\Delta t_{\s P}=6$\,s (brown curves) and $\Delta t_{\s P}=60$\,s (grey curves). We can note that, for $\Delta t_{\s P}=60$\,s, Hydra is trapped in the 6/1 MMR,  with the critical angle $\phi_{6/1}=6\lambda-\lambda_{\s C}-\varpi_{\s C}-4\varpi$ {librating} around $0^{\circ}$. Moreover, the ratio $n_{\s C}/n$ is slightly smaller than 6 for this specific trapping. In the case with $\Delta t_{\s P}=6$\,s, all critical angles of the $6/1$ MMR are circulating.

{As in the case of Kerberos, all initial conditions result in stable motion of the test moons, with $e<0.05$, indicating good agreement with the current orbit of Hydra.}


\subsection{Discussion of moons behaviour and proximity to MMRs}

Analysing the parameter's values in Table \ref{tab-results}, we realise that it is unlikely to collect a specific set of the parameters and initial conditions resulting in the final configuration of the all four satellites consistent with the current ones. {This fact is mainly due to the difficulty to find Styx in stable configuration.}

Since the  determination of the satellite's current orbits is affected by uncertainties, {it is not clear whether they are or not captured in the MMRs.} Thus, for sake of completeness, we show in Table \ref{tab-captures} the values of the parameters, which provide the final moon's orbits trapped in the nearest MMR with Charon, with at least one of the associated critical angles librating. It is remarkable that the motion of Styx in the 3/1 MMR is always unstable, in this way confirming our suggestion that the 3/1 MMR delimits the domain of stable motion in the \PC binary.

{For sake of completeness, we compare our results with those previously reported by \citet[][]{Woo+2018} assuming the tidal model of constant $\Delta t$. Their model considered $A=10$ and $A=40$ and initial $e_C=0$ and $e_C=0.2$. For large $A$ and initial $e_C=0$, all test particles survives the entire simulation (see their Table 3, last row). Moreover, the particles nearby the MMRs with Charon have final values of eccentricities larger than the current ones. 
On the other hand, for small and large $A$ and initial $e_C=0.2$, the final orbits do not match the current ones.
These results are in good agreement with ours, however, \citet{Woo+2018} only considered fixed values of $\Delta t=600$ s, $\Omega_P/n=5.65$, and initial distance for Charon $a_C=4 R_P$, while our study present a wider range of initial parameters and tidal dissipation.}


\begin{table}
\caption{List of the parameters resulting in captures in the MMRs with Charon. Styx's simulations did not produce any captures in the MMRs.}
\begin{center}
\begin{tabular}{|c|c|c|c|c}
\hline
Parameter & Nix & Kerberos & Hydra \\ 
\hline
$a_{\s C}/R_{\s P}$     &  3, 5, 6 & 3 & 3, 5, 6 \\
\hline
$e_{\s C}$              & -  & 0, 0.001 & 0, 0.001  \\
\hline
$\Omega_{\s C}/n_{\s C}$&  4 & 0.5, 1,5, 3, 4 & 0.5, 1,5, 3, 4 \\
\hline
$A$                     & -  & $0.01, 0.1, 1, 100$ & $0.01, 0.1, 1, 100$\\
\hline
$\Delta t_{\s P}$\,(s)  & 60 &      60             & 60 \\
\hline 
\end{tabular}\label{tab-captures}
\end{center}
\end{table}


\subsection{Survival time of the small moons under $J_2$ evolution} \label{sec.constantmass}

In order to refine our model, we introduce an additional force due to the oblateness of Pluto. 
The Pluto's polar oblateness, parameterised by the zonal harmonic $J_2$, is modelled as \citep{Chen+2014b}
\begin{equation}\label{eq.j2}
J_2=\frac{k_{\s f P}R_{\s P}^3\Omega_{\s P}^2}{3Gm_{\s P}},
\end{equation}
where $k_{\rm \;f\;P}$ is the second order fluid Love number of Pluto and $\Omega_{\s P}$ is the angular rotation velocity of Pluto. {Recently, \citet{Correia+2020} derived from the shape of Pluto a static value of $J_2=6\times 10^{-5}$}. In this work, we use both formulations adding them to the model described in Sect.\,\ref{sec.tidalmodel}.

{We apply the model to study the behaviour of a ring composed of 300 particles, each one of the very small mass of $1.49 \times 10^{15}$\,kg (thus, we can neglect their mutual perturbations and contributions in the total angular momentum of the system), at different positions chosen from the interval of $10R_{\s P} <a < 55R_{\s P}$ of the baricentric semimajor axis. We use the constant $\Delta t$-model, $A=10$ and the different initial values of the Charon's eccentricity, $e_C$. It is worth noting that the value of $e_C$ constrains $\Omega_{\s P}$ (equivalently, the initial rotation period of Charon). For example, starting at $a_C=4 R_P$, with $e_C=0.001$ and $\Omega_{\s C}/n_{\s C}$=2 (i.e. $P_{\s C}=0.38136$\,d) and using Eq.\,(\ref{eq.angmot}), we calculate the corresponding initial rotation period of Pluto as being $P_0=0.13858$\,d ($P_0=2\pi /\Omega_{\s P}$)}. 

  \begin{figure}
  \centering
\mbox{\includegraphics[width=\columnwidth]{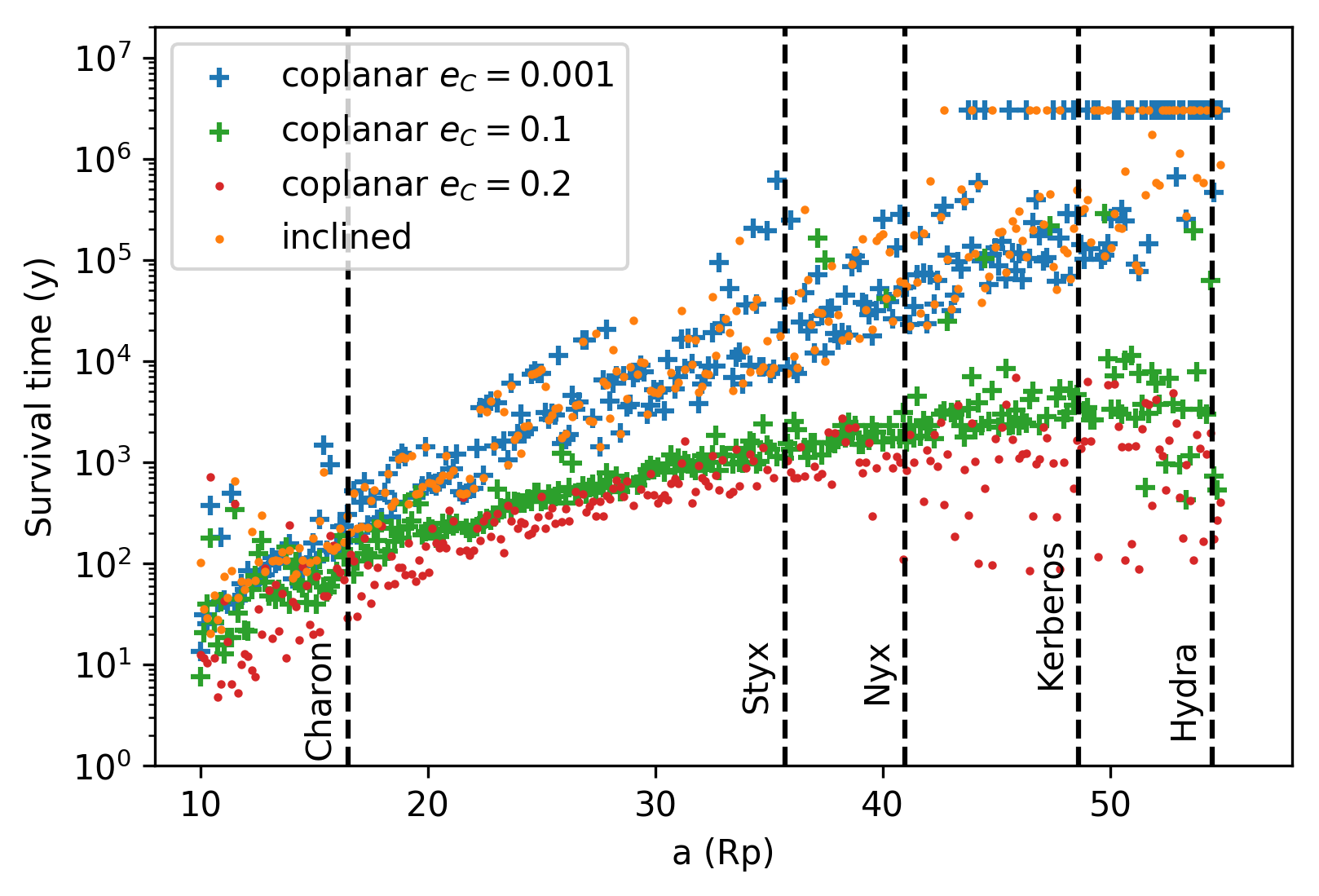}} \\
\caption{Survival times of the small particles, for the different discs extending from $10 {R_{\s P}}$  to $55 {R_{\s P}}$,  during the tidal expansion of the \PC binary, with the dissipation parameter $A=10$, under the effect of the {evolving} zonal harmonic $J_2$. The colour symbols are used to identify the particles from the different discs (see text).}
  \label{fig-survival-J2evol}
  \vspace{-1em}
  \end{figure}
  

  \begin{figure*}
  \centering
\mbox{\includegraphics[width=0.95\textwidth]{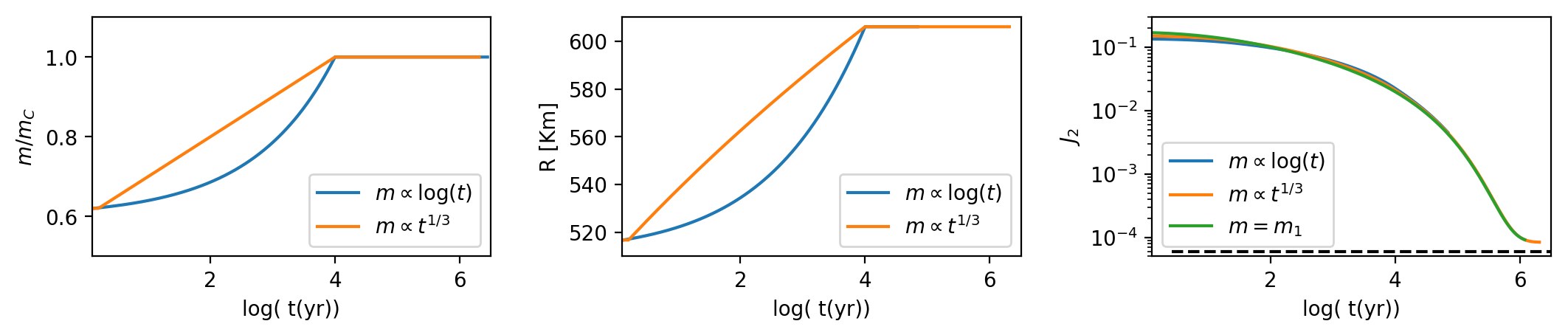}} \\

\caption{Time evolution of the Charon physical properties {during the mass accretion}: {the mass, the radius, and the zonal harmonic $J_2$, from the left to the right panels, respectively}. Colours represent the results obtained from the different accretion models. Dashed horizontal line (right panel) present the constant  $J_2$ obtained in \citet{Correia+2020}.}
  \label{fig-mrj2}
  \vspace{-1em}
  \end{figure*}

We work with two kinds of the disc of particles, one is coplanar with the orbital plane of the \PC binary, and other is inclined, with the angle randomly chosen from the interval $-0.01^\circ < i <0.01^\circ$. In the coplanar disc, the particles start on the nearly circular orbits, with eccentricities uniformly distributed in the interval $0.0<e<10^{-3}$. For the Charon's initial configuration, we consider $a_C=4 R_P$ and three values of the eccentricity $e_C$, namely, $0.001$, $0.1$ and $0.2$. In the case of the inclined disc, the particles remain on the nearly circular orbits, while $a_C= 4 R_P$ and $e_C=0.001$. In both cases, we set random initial conditions for the angular variables of the Charon's and particle's orbits. The system composed of the \PC binary and the ring of particles was integrated over $3 \times 10^6$\,yr (approximately $150 \times 10^6$  periods of the binary)\footnote{According to \cite{Chen+2014a, Correia+2020}, the \PC binary acquires its equilibrium current position after $(2 - 4) \times 10^6$\,yr.}. 
 
The case of the initially inclined disc of particles is considered due to the recently published results \citep{Cuk.2020}, which demonstrate that tidal evolution of particles at low inclinations, can induce resonant trapping and the excitation of the particle's inclination, that is incompatible with the observed low inclinations of the moons in the \PC system. The tests with the initially eccentric orbit of Charon are inspired on the results of the smooth particle hydrodynamic simulations reported in \citet[][]{Canup2005,Canup+2011}. 

Figure \ref{fig-survival-J2evol} shows the survival times of the particles from the different discs. The results were obtained considering the zonal harmonic $J_2$, which evolved as the \PC binary  tidally expands, according to Eq.\,(\ref{eq.j2}). We have also investigated the behaviour of the disc considering the cases of $J_2$=0 and $J_2=6 \times 10^{-5}$, this last value derived from \citet{Correia+2020}; the results obtained are discussed in Appendix.  We note in Fig.\,\ref{fig-survival-J2evol} that the particles could survive the tidal expansion of the \PC binary ({that is, remain in the \PC system after $3 \times 10^6$\,yr}) only when Charon starts on the nearly circular orbit ($e_C=0.001$), {and only in the region beyond the Nix's orbit}. Also,  there are no significant differences between the coplanar and low inclined discs (blue crosses and orange dots, respectively). Contrarily, in the case of the initially eccentric Charon's orbit, all particles (green crosses and red dots) are ejected from the system in less than $10^3$\,yr, due to capture in the MMRs and subsequent excitation of the eccentricities, as described in \citealt{Chen+2014a, Kenyon+BromleyIII}. 

For the quasi-circular Charon's orbit, the low order 3/1 and 4/1 MMRs, at the positions of Styx and Nix, respectively, are responsible for the rapid (less then $10^4$\,yr) ejection of the particles. In this case, the survivors are observed only beyond the Nix's orbit. Moreover, even further, in the regions around current positions of Kerberos and Hydra (5/1 and 6/1 MMRs, respectively), the most of the fictitious moons do not survive more than $10^5$\,yr. It is clear that these times are incompatible with the hypothesis of primordial moons after the system settled into its present configuration through an impact on Charon \citep{Bromley+2020arXiv200613901B}.

{It is worth observing that, applying the simplified model, which considers the constant value of $J_2$ (zero or $J_2=6 \times 10^{-5}$), we obtain that the  particles starting on the nearly circular and nearly planar orbits at the current positions of Styx and Nix, can survive the tidal expansion of the \PC binary (for details, see Appendix). However, in the more realistic model, which accounts for the zonal harmonic $J_2$ evolving during the expansion, the additional perturbations  destroy the stability of the particles in the range of $10R_P<a<55R_P$.  Moreover, no particles remain at the current positions of Styx and Nix. We repeat these experiments with a higher dissipation ratio, {$A$=20}, and obtain qualitatively the same result.}

  \begin{figure*}
  \centering
\mbox{\includegraphics[width=0.85\textwidth]{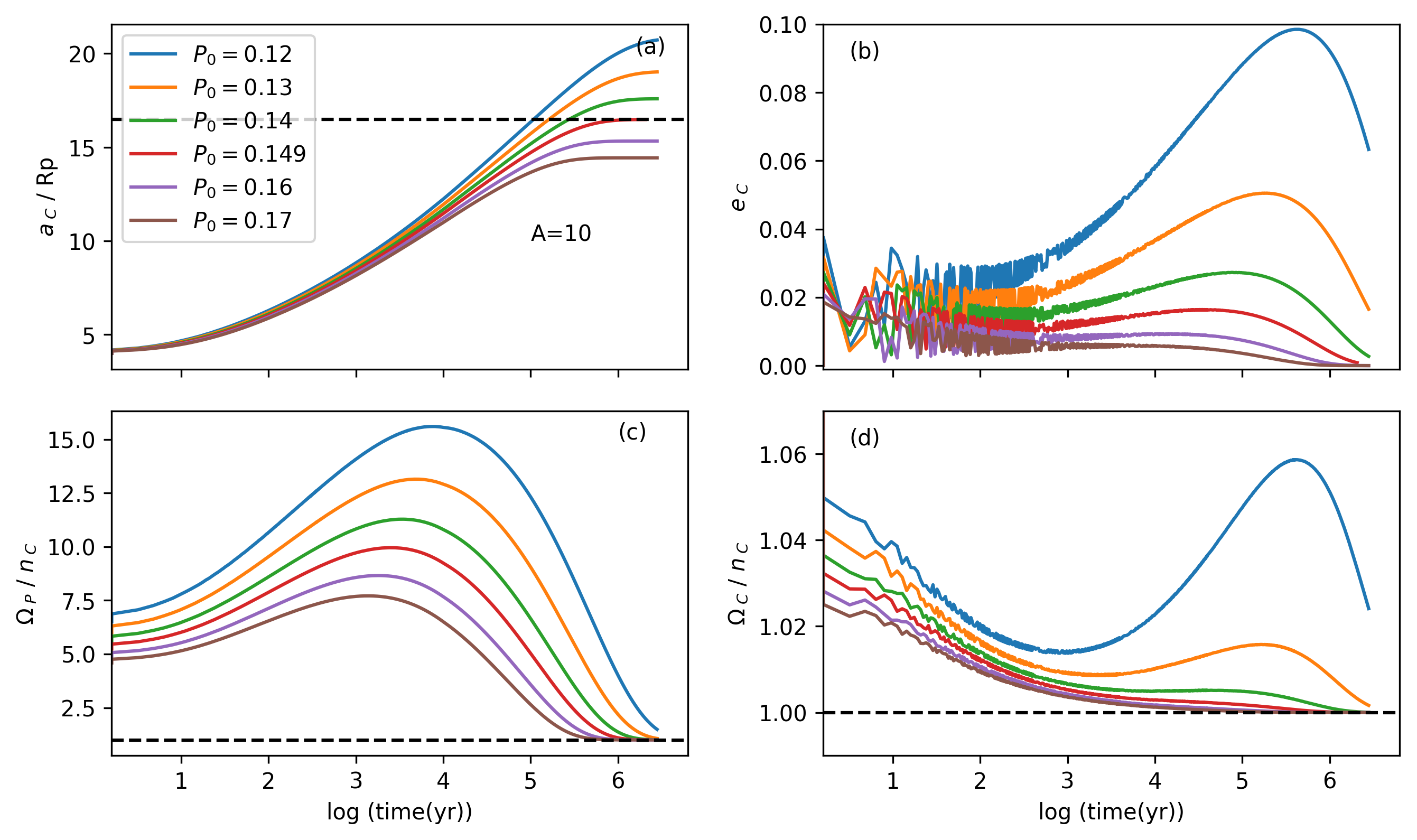}} \\
\caption{Tidal evolution of the \PC binary, for $A=10$ and logarithmic mass accretion of Charon during the first $10^4$\,yr, starting with the initial eccentricity $e_C=0.001$. Panels show the orbital semimajor axis $a_C$ (panel {a}), in units of $R_{\s P}$, the orbital eccentricity $e_C$ (panel {b}), and the spin angular velocities of Pluto $\Omega_{P}$ (panel {c}) and Charon $\Omega_{C}$ (panel {d}), in units of the mean motion $n_C$. Each colour represent a different initial $P_0$, in units of days.}
  \label{fig-j2A10}
  \vspace{-1em}
  \end{figure*}
  
  \section{Accreting Charon's mass}\label{sec.massgrowth}
  
In this section, we introduce a scenario, which considers the mass accretion of proto-Charon from the disc of debris left soon after the giant impact. The growth of Charon occurs simultaneously with the tidal expansion of its orbit around Pluto and could affect the behaviour of the small circumbinary moons described in the previous sections.

First, we analyse the evolution of the expanding binary with the growing Charon in Sect.\,\ref{sec.pcaccrenting}. Then, we introduce an additional small moon (either Styx or Nix),  to assess the  parameters of the tidal evolution in Sect.\,\ref{sec.pcpaccrenting}. Finally, we simulate the behaviour of the disc of particles in the \PC system in Sect.\,\ref{sec.ringaccreting}. {In this section we choose the plane of the Charon's orbit as a reference plane. When the planar case is considered, the initial inclinations do no shown.}
  
\subsection{Tidal evolution of the \PC binary with the accreting Charon's mass} \label{sec.pcaccrenting}

We start considering the evolution of the \PC pair described in Sect.\,\ref{sec.tidalmodel}, now including the accretion of Charon from the disc of debris. The duration of the accretion process is fixed arbitrarily at $10^4$\,yr since a giant impact, starting with the Charon's mass of $60 \% $ of its current value, that is in accordance with the results in \citet{Canup2005} and \citet{Kokubo+2000}. Following \citet{Kokubo+2000}, we adopt a logarithmic law for the mass accretion, but we also consider an oligarchic growth defined as $m \propto t^{1/3}$. The evolution of the increasing mass of Charon is shown on the left panel in Fig.\,\ref{fig-mrj2}, for both models. Assuming Charon as an homogeneous sphere, we calculate its radius for the constant density of $\rho = 1.70 \text{ g/cm}^{3}$,  and show its time evolution on the middle panel in Fig.\,\ref{fig-mrj2}. Finally, the right panel shows the time evolution of the zonal harmonic $J_2$ of the Pluto's oblateness during the tidal migration of the growing Charon. According to Eq.\,(\ref{eq.j2}), $J_2$ is a function of the orbital spin of Pluto $\Omega_{\s P}$, whose starting value is chosen as described below. For the chosen $\Omega_{\s P}$, $J_2$ starts at $\sim 0.17$ and reaches $8 \times 10^{-5}$ at the end of the migration, for both logarithmic and oligarchic laws of the mass accretion. We note that both models represent almost the same evolution of $J_2$. We also show in Fig.\,\ref{fig-mrj2}\,right panel the constant $J_2$ of $6 \times 10^{-5}$ derived in \citet{Correia+2020}.

It is worth emphasising that, in the growing mass scenario, the total angular momentum of the \PC binary given in Eq.\,(\ref{eq.angmot}) is no longer conserved. Thus, we need to obtain the initial value of the Pluto's spin $\Omega_{\s P}$ and the dissipation ratio $A$, in such a way that both can guarantee the current double synchronous configuration of the binary, with orbital parameters compatible with those from Table \ref{tab-init}.

   \begin{figure}
  \centering
\mbox{\includegraphics[width=0.9\columnwidth]{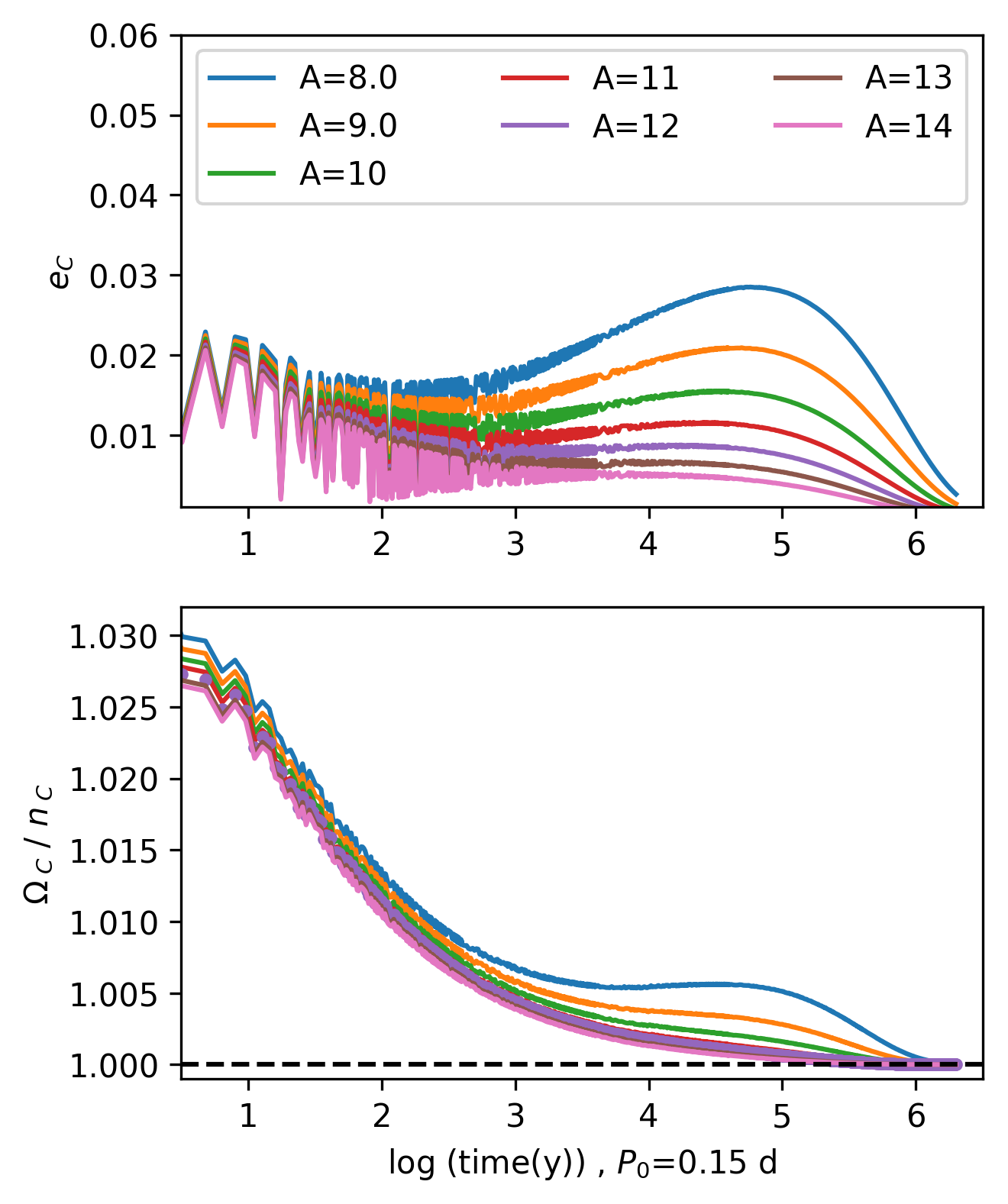}} \\
\caption{Time evolution of the \PC binary, eccentricity (top panel) and $\Omega_{\s C}/n$ (bottom panel), for the fixed initial $P_0=0.149$\,d and different values of $A$ represented with different colours.}
  \label{fig-j2Aii}
  \vspace{-1em}
  \end{figure}

    \begin{figure}
  \centering
    \mbox{\includegraphics[width=0.9\columnwidth]{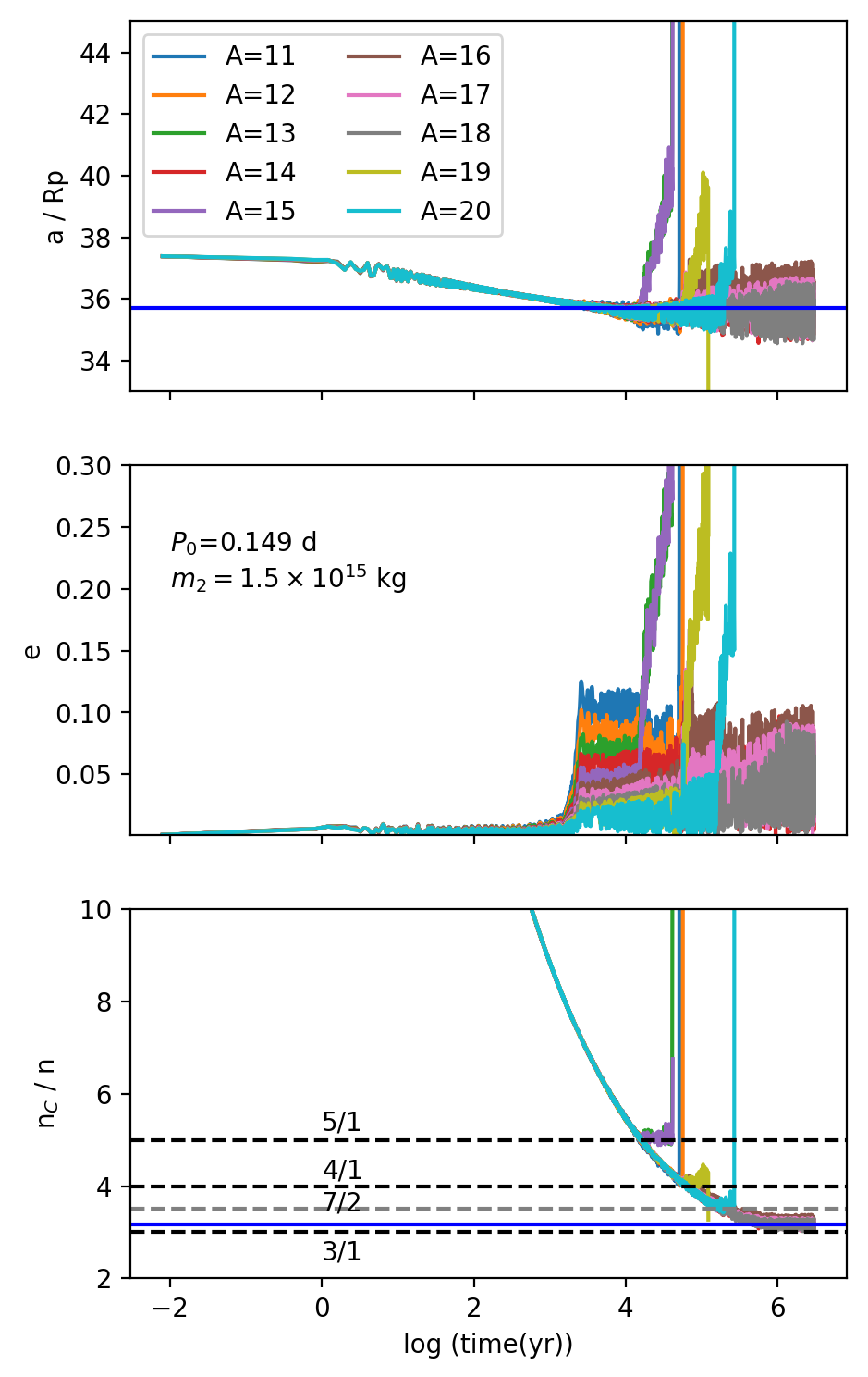}}
\caption{Time evolution of the orbital elements of Styx, with the mass $m_S=1.49 \times 10^{15}$\,kg,  for the different dissipation values of $A$. {Top-panel shows the Styx's semimajor axis, the middle-panel shows the eccentricity and the bottom panel shows $n_C/n$. The continuous horizontal line on the top panel indicates the current Styx's position, at $a/R_p=35.7$, while dashed lines on the bottom panel identify 3/1, 4/1 and 5/1 MMRs.} }
  \label{fig-Styx1e15}
  \vspace{-1em}
  \end{figure}

    \begin{figure}
  \centering
\mbox{\includegraphics[width=0.9\columnwidth]{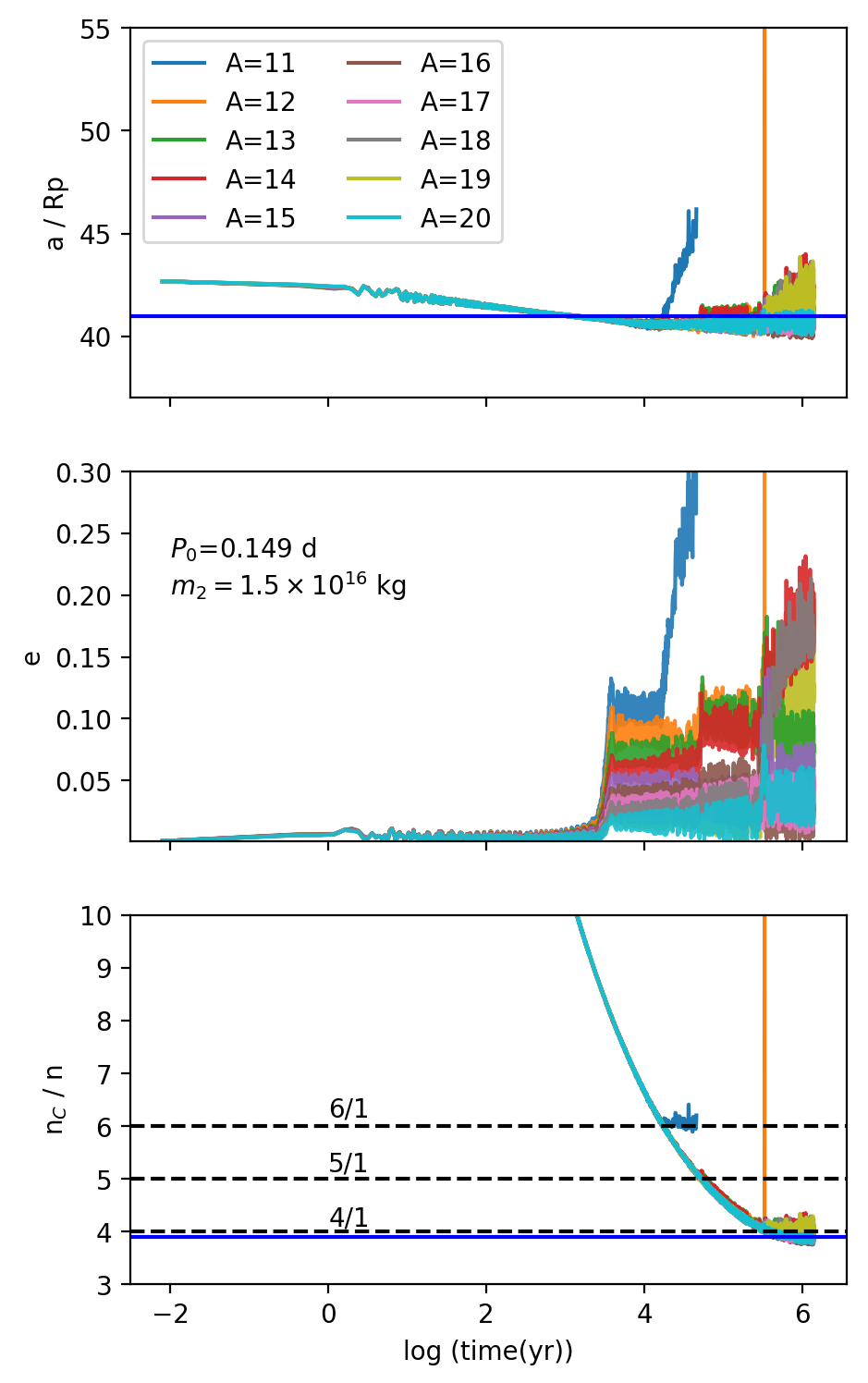}}
\\
\caption{Same as Fig.\,\ref{fig-Styx1e15}, except for Nix, with the mass $m_N=1.5 \times 10^{16}$\,kg at the position $a/R_p=40.98$ (continuous horizontal line).}
  \label{fig-Nix}
  \vspace{-1em}
  \end{figure}
  
Using a logarithmic law for the Charon's accretion  and fixing $A=10$ and $e_{\s C}=0.001$, we vary the initial rotation period of Pluto, $P_0=2\pi /\Omega_{\s P}$, in the range from 0.12\,d to 0.21\,d, in order to determine the value that assures, at the end of the migration, the current configuration of the \PC binary. We choose the low initial eccentricity of Charon due to constraints already described in Sect.\,\ref{sec.constantmass}. We simulate the tidal evolution of the \PC binary and show the results in Fig.\,\ref{fig-j2A10}, where the time evolution of the Charon's semimajor axis $a_{\s C}$ (in units of $R_{\s P}$) is presented on the panel {(a)}. The horizontal dashed line, which shows the final state of the current system, corresponds to the initial value $P_0=0.149$\,d. The smaller initial values of $P_0$ lead to the wider \PC binaries, while the larger values of $P_0$ lead to more compact systems, comparing to the current \PC configuration. The eccentricity of the binary is more efficiently damped for higher  initial values of $P_0$, as shown on the panel {(b)} in Fig.\,\ref{fig-j2A10}. The panels {(c)} and {(d)} show the evolution of the angular velocities of Pluto and Charon, $\Omega_{\s P}$ and $\Omega_{\s C}$ (in units of the mean motion $n_{\s C}$), respectively; it is clear that the system reaches the double synchronous rotation more rapidly for the higher values of $P_0$.    

The time evolution of the accreting system for the different values of the dissipation parameter $A$ is presented in Fig.\,\ref{fig-j2Aii}. The values of $A$ are chosen from the range $[8-21]$, while the initial  rotation period of Pluto and the initial Charon's eccentricity are fixed at $P_0=0.149$\,d and $e_C=0.001$, respectively.  The Charon's eccentricity, shown on the top panel in Fig.\,\ref{fig-j2Aii}, is damped more efficiently as $A$ increases. The evolution of $\Omega_{\s C}$ (in units of the mean motion $n_{\s C}$) is shown on the bottom panel; we note that the synchronous state of Charon is more quickly attained for higher values of the  dissipation parameter $A$. The variations of the semimajor axis of Charon and the rotation of Pluto $\Omega_{\s P}$  with the increasing $A$-value are insignificant and are not shown in Fig.\,\ref{fig-j2Aii}.

The tests described above allow us to define the parameters' values, which are compatible with the current configuration of the \PC binary, being $P_0=0.149$\,d and $A$ for the interval $[8-21]$. The tests, done with the oligarchic mass accretion,  indicate the initial value of $P_0=0.157$\,d; however, the corresponding simulations have shown the results, which were qualitatively similar to those obtained with the logarithmic mass growth model.

\subsection{\PC binary and a small moon} \label{sec.pcpaccrenting}

In this section, the model is expanded adding a small moon, with the mass and orbital elements of either Styx or Nix. Indeed, as described in the previous section, the behaviour of these two satellites is significantly affected by the passages through the strong 4/1  and 5/1 MMRs. We explore the parameter space of the problem defined by $A$ and $m_i$, being $m_i$ the mass of the small moon. For the mass, we choose the extreme values of the masses of Styx and Nix, given by the uncertainties reported in  \citet{Brozovic+2015} and \citet{Showalter+2015}. We set the initial eccentricity of the small moon at $e=0.001$, the semimajor axis at either $a = 44413$\,km ($a \sim 37.4 R_P$) or  $a = 50690$\,km ($a \sim 42.7 R_P$), for the Styx and Nyx clones, respectively, and the angular elements equal to zero. As shown below, these initial values of $a$ are chosen to obtain the current positions of the small moons. 
All simulations in this section were done with the accretion time of $10^4$ yr.


\begin{figure}
  \centering
\mbox{\includegraphics[width=1\columnwidth]{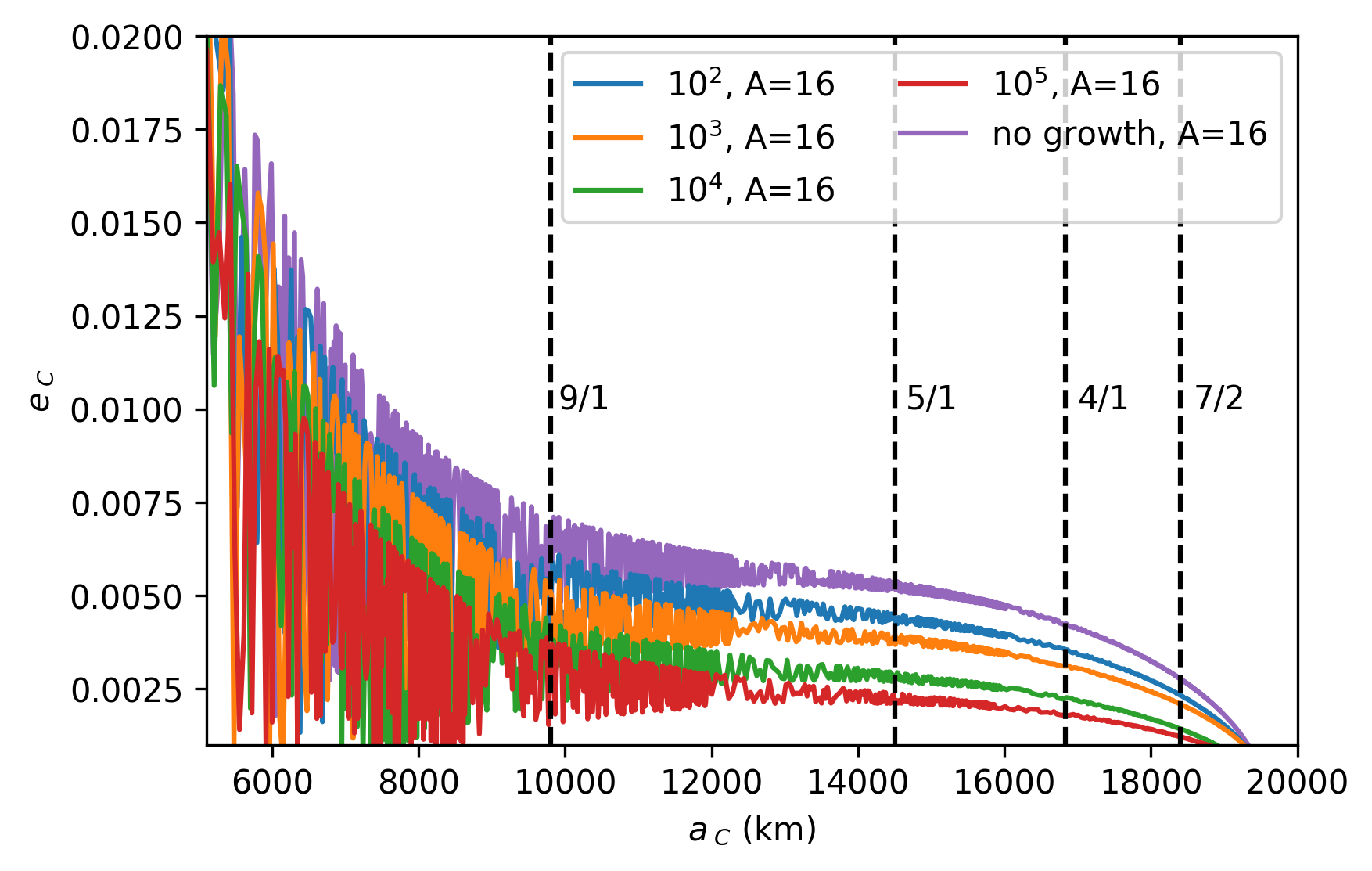}}
\caption{Eccentricity of the growing Charon during the tidal expansion of the Charon's orbit. Different colours are used to present the different values of the accretion time. The vertical dashed lines show the nominal locations of the main MMRs originated by the migrating Charon at the Styx's current position; each MMR is identified by the corresponding label.}
 \vspace{-1em}
   \label{fig:ac-ec-growth}
  \end{figure}
  
\begin{figure}
  \centering
\mbox{\includegraphics[width=0.95\columnwidth]{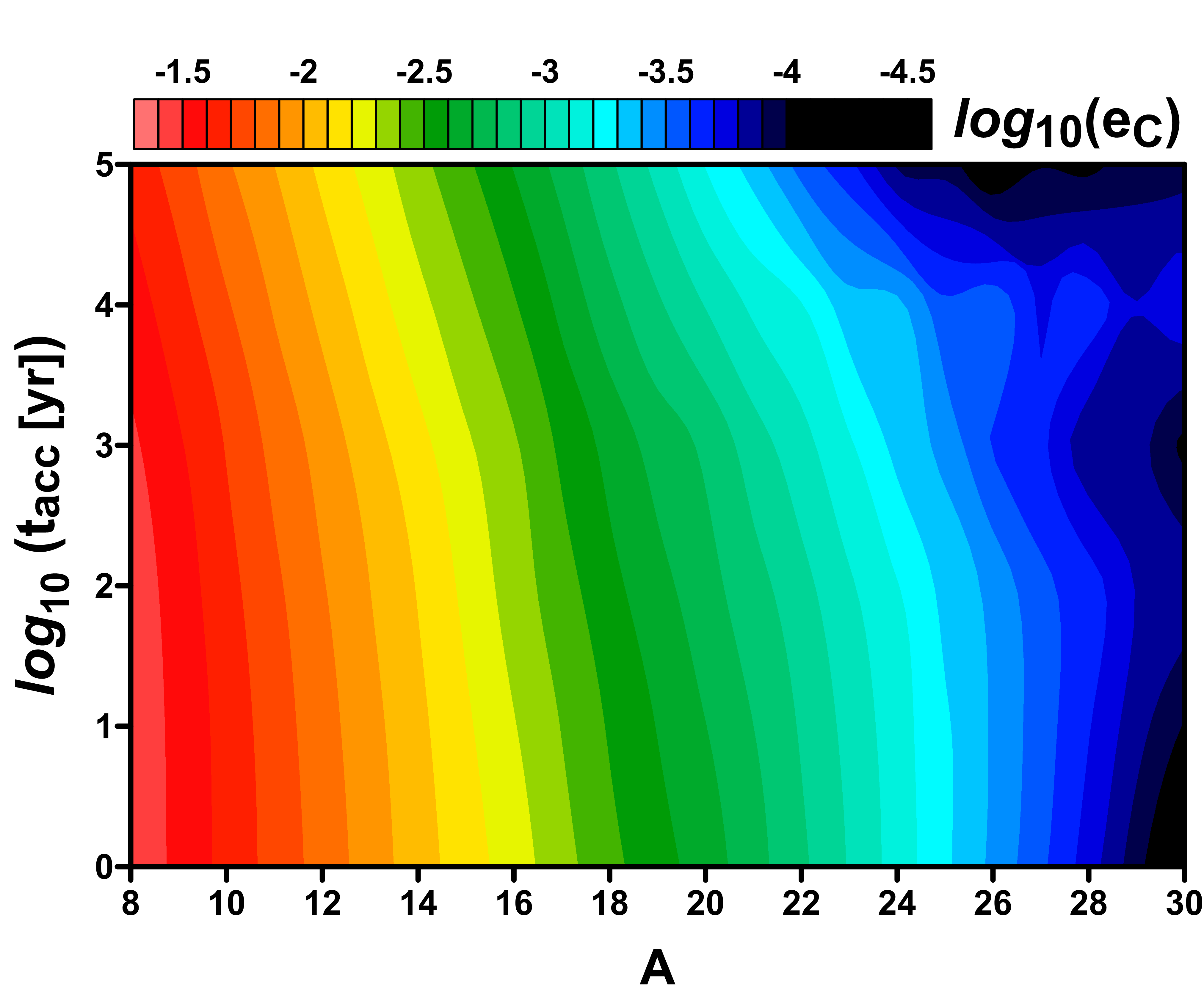}}
\caption{Parametric $A$--$t_{acc}$ plane  shows, in colour scale,  the values of the Charon's eccentricity, $e_{\s C}$ (in logarithmic scale), at the instant, in which the 5/1 MMR crosses the Styx's current location. }
 \vspace{-1em}
   \label{fig:A-t-plane}
  \end{figure}

We start analysing the behaviour of the Styx clones during the tidal evolution of the growing Charon. {We consider first the Styx's mass of $1.49 \times 10^{15}$\,kg, which corresponds to the $1\sigma$ estimate given in \citet{Brozovic+2015}. }
Figure \ref{fig-Styx1e15} shows the time evolution of the Styx's orbital elements obtained for the values of the dissipation coefficient $A$, ranging from 11 to 20; the elements are the semimajor axis (top panel) and the eccentricity (middle panel). We note that the test moon smoothly migrates inward during the first $\sim 10^4$\,yr {that is related to the increasing Charon's mass}. After the satellite reaches its current position (continuous horizontal line), it evolves with a nearly constant semimajor axis, up to encounter to some important MMRs with Charon. 

The interaction of the moon with the MMRs is generally accompanied by the excitation of the moon's eccentricity. We can observe this phenomenon on the {middle panel} in Fig.\,\ref{fig-Styx1e15}, which shows the time evolution of the moon's eccentricity during the migration of the growing Charon. When the Charon's mass grows up to 90\% of its current value (in about $3\times 10^3$ yr), the moon's eccentricity is excited due to the approximation to the resonance domain, in this case, of the 9/1 MMR. From that event, the eccentricity can suffer consecutive excitations during the passages of the moon through  the resonances, up to be even ejected from the system, depending on the value of $A$. The current configuration of Styx is acquired {for around 30\% of the tested values of $A$, namely $16$, $17$, and $18$}. {As observed on the bottom panel in Fig\,\ref{fig-Styx1e15}, there are two ejections due to the 5/1 MMR, four due to the 4/1 MMR, and one is associated to the 7/2 MMR.}

\begin{figure}
  \centering
\mbox{\includegraphics[width=1\columnwidth]{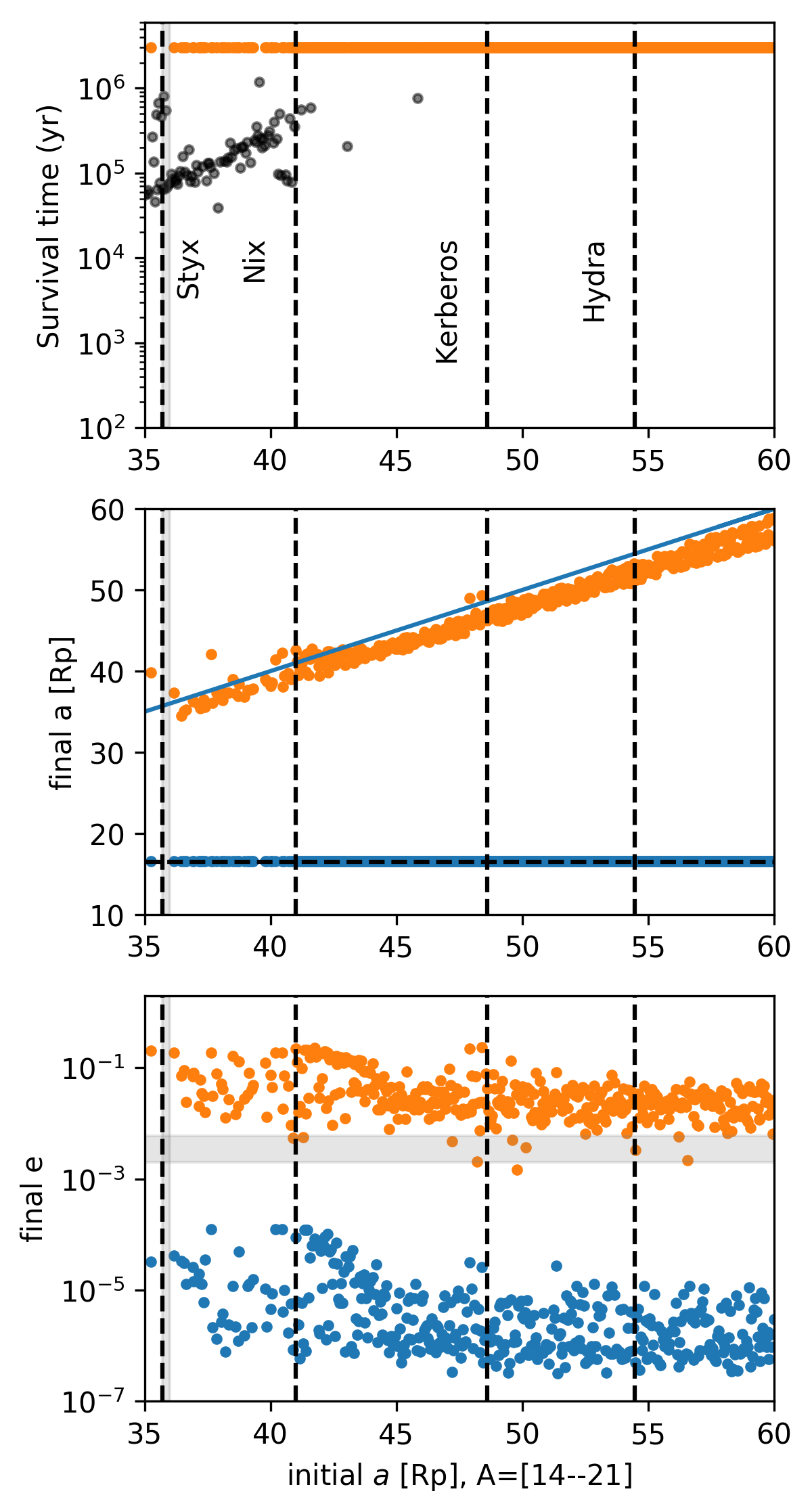}}
\caption{Evolution of the ring of particles each of the mass of $4.5 \times 10^{16}$\,kg, for $A$ from the range [14--21]. Top panel shows survival times for particles, middle panel the final semimajor axis of the small moons and Charon, and bottom panel the final eccentricity of Charon and particles, respectively. Mass growth of Charon is set in $10^4$ yr. Blue colour is for orbital parameters of Charon while orange for the particles. Top panel also present black points identifying those particles that escape while \PC binary evolve. Shaded grey region indicates the range of values according to diverse authors (see Table \ref{tab-init}).}
 \vspace{-1em}
   \label{fig-Particles}
  \end{figure}

\begin{figure}
  \centering
\mbox{\includegraphics[width=1\columnwidth]{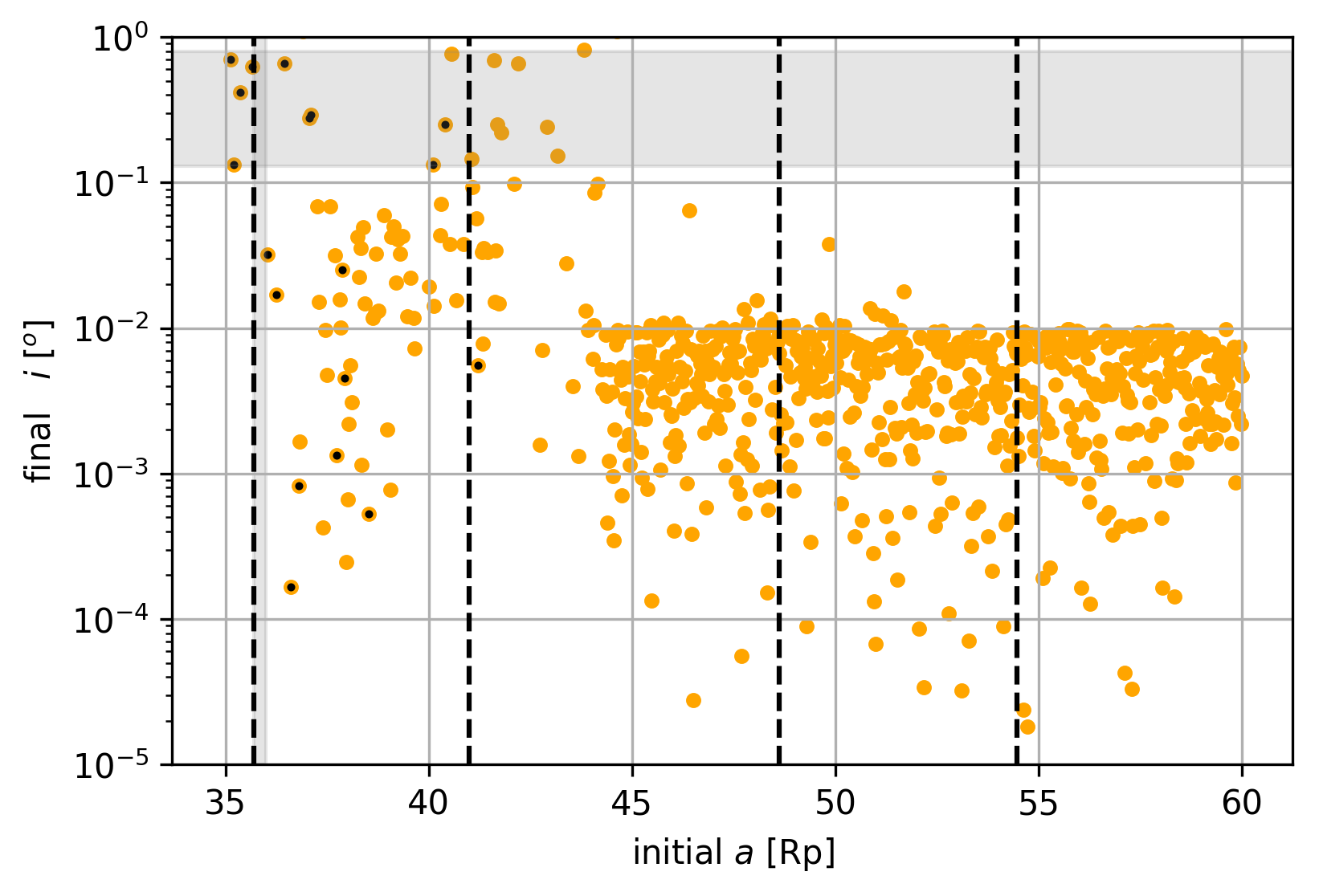}}
\caption{Distribution of inclination for the disk of particles after the dual synchronisation of \PC system with mass growth. The orbits that do not survive the entire integration time are identified with black dots. Shaded grey region indicates the range of values according to diverse authors (see Table \ref{tab-init}).}
  \label{fig-Particlesinc}
 \vspace{-1em}
  \end{figure}

We repeat the simulations described above, but now analysing the behaviour of the Nix clones during the tidal expansion of the accreting Charon. {We consider the value of the Nix's mass of $1.5 \times 10^{16}$\,kg, which is smaller from the mass range obtained in the orbital fits reported in \citet{Brozovic+2015} and \citet{Showalter+2015}}. The results obtained for Nix are shown in Fig.\,\ref{fig-Nix}, which is similar to Fig.\,\ref{fig-Styx1e15}, but, in this case, {we observe fewer ejections: one from the 6/1 MMR, and other from the 4/1 MMR (see bottom panel). We can observe on the middle panel of Fig.\,\ref{fig-Nix} that the particles experience consecutive eccentricity excitations during the passages through the MMRs, resulting in the higher dispersion of the final values, when compared to the case of the Styx's clones}. 

{We redid the simulations described above for the Styx and Nix clones, considering now the larger masses, $1.5 \times 10^{16}$\,kg and $4.5 \times 10^{16}$\,kg, respectively. We could not observe significant differences comparing the new results to those obtained for the smaller masses of the test moons. Thus, we conclude that, in the range of values considered in this paper, the masses of the moons are not an important factor to achieve favourable configurations of the satellite system. On the other hand, the simulations done with the values of the dissipation coefficient $A$  in the range $11-20$, did not allow definitive conclusions on the influence of this parameter on the moon's behaviour. In the next section, we return to explore this issue. } 

\subsection{Dependence on the accretion time}
\label{sec:t-accretion}

The accretion time of Charon, $t_{acc}$, is an important parameter, since it affects the time evolution of the Charon's orbital elements, particularly, the Charon's eccentricity. As shown in Sect.\,\ref{PC+moon}, the ejection of a small circumbinary moon during the passage through one of the main MMRs is most likely for the higher values of the Charon eccentricity. On the other hand, the rate of the Charon's mass accretion determines the damping rate of its eccentricity. Indeed, Fig.\,\ref{fig:ac-ec-growth}  shows the evolution of $e_{\s C}$ during the tidal expansion of the Charon's orbit. We use the different colours to identify the tested values of the accretion time, namely $10^2$\,yr, $10^3$\,yr, $10^4$\,yr and $10^5$\,yr; the curve corresponding to the simulation with the 'no-growing' Charon, starting  with its current mass, is also shown in Fig.\,\ref{fig:ac-ec-growth}. All simulations were performed with the initial value of $e_{\s C} = 0.001$ and the $A$-value fixed at 16. It is worth warning that the accretion rate determines the initial value of the Pluto's rotation period $P_0$;  for the tested times of $10^2$\,yr, $10^3$\,yr, $10^4$\,yr and $10^5$\,yr, we obtain the optimal values as $P_0=0.1400$\,d, $P_0=0.1437$\,d, $P_0=0.149$\,d, and $P_0=0.1548$\,d, respectively (see Sect.\,\ref{sec.pcaccrenting} for details). 

Figure \ref{fig:ac-ec-growth} shows that the damping of the Charon's eccentricity is more efficient for the increasing $t_{acc}$.  Indeed, we can clearly observe that the Charon's eccentricity systematically increases with the decreasing accretion time, and reaches its maximum value in the case of the non-accreting process. The vertical dashed lines mark the main MMRs and allow us to evaluate $e_{\rm C}$,  when one of these MMRs crosses the current position  of Styx. Hence, during the passages of the small moons through the MMRs, their motions will be  more excited by Charon accreting the mass more rapidly.  This fact would explain why we could not obtain successful results simulating the evolution of the \PC system without considering the accretion of the Charon's mass (see Sect.\,\ref{sec.constantmass}).  

As shown in the previous sections, another important parameter of the problem is the dissipation coefficient  $A$. In order to link information on the dissipation parameter $A$ to information on the accretion times, we prepare the parametric $A$--$t_{acc}$ plane, which is shown in Fig.\,\ref{fig:A-t-plane}. For a given point on the plane, we show the value of $e_{\rm C}$ (in logarithmic scale) at the instant, in which the 5/1 MMR originated by migrating Charon, crosses the current position of Styx. For this, we use the colour scale palette shown on the top of the figure. We note that the zero-value on the y-axis corresponds to the non-accreting Charon's mass simulations.
The hot-red domains are associated with the highest $e_{\rm C}$-values, while the dark-blue domains are related to the lowest values. Observing the plane, we can conclude that the efficient damping of $e_{\rm C}$ during the tidal expansion occurs for  higher values of diffusion parameter and the slower rate of accretion that favours the survival of the small satellites. 
 
\subsection{Ring of particles in the accreting \PC binary}
\label{sec.ringaccreting}

In this section, we consider the behaviour of the circumbinary thin disc of 500 particles during the tidal expansion of the accreting \PC binary. Following  \citet{Kenyon+BromleyIII}, we first choose the disc, which is initially extended from $34 \; R_p$ to $60 \; R_p$, with the eccentricities of the particles uniformly distributed between $0.5 \times 10^{-3}$ and $1.5 \times 10^{-3}$,  no inclinations with respect to the orbital plane of the \PC binary, and randomly chosen angular elements. 
{We work with the particles with the mass of either $1.49\times 10^{15}$\,kg  or $4.5 \times 10^{16}$\,kg}; these values are still smaller than the mass of the \PC binary; thus, the particle's interactions can be neglected. {Also, we select random values of the dissipation parameter $A$ in the interval of $ [14 - 20] $, because these values favour the survival of the Styx and Nix clones, as described in the previous section.} Finally, the time of the accretion of the Charon's mass is $10^4$\,yr.

Figure \ref{fig-Particles} shows the results obtained for the disc of particles with the mass of $4.5 \times 10^{16}$\,kg, as a function of the initial positions of the particles. The top panel shows the survival times of the test particles: orange dots represent survivors ({the particles that remain close to initial configurations during the integration time of $3 \times 10^6$\,yr), while black dots represent the particles that are ejected from the system during the integration.}
In the regions beyond the Nix position, {almost all} particles survive {the period of the tidal expansion of the accreting \PC binary}{}, while the region between Styx and Nix presents more instabilities (black dots in top panel). In the middle panel of Fig.\,\ref{fig-Particles}, we plot the final positions of the particles (orange) and Charon (blue). The final positions of the particles are generally closer to the binary than initial conditions (the points lie slightly below the line $a_i=a_f$, {plotted as a continue blue line}); {this is a consequence of the increasing mass of the binary}. 

Figure \ref{fig-Particles}\,bottom shows the final eccentricities of the \PC binary (blue dots) and the particles (orange dots). For the particles with {initial $a$} chosen in the range of $ 34 \; R_p - 45 \; R_p $, the eccentricities are slightly higher than those from the range $45  \; R_p - 60 \; R_p $. The shaded region represents the extreme values of the eccentricities of the known moons, according to Table \ref{tab.Pluto}; the output of the several simulations end in this shaded region. The \PC binary is almost circular, with $e_C \sim 10^{-5}$, that is compatible with its orbital fits \citep{Brozovic+2015, Showalter+2015}. {We do not find any preference for the $A$ values in the range of $ [14 - 20]$ in the region, which is interior of the Nix position; around 35\% of particles survive in this region. We repeated the  experiment choosing the $A$ values in the range of $ [9 - 13] $ and found that  99\% of clones located inside the Nix's orbit were ejected. }

Several works have shown that the particles starting with low initial inclinations, could be captured in a secular resonance, with a consequent excitation of their inclinations \citep[e.g.,][]{Cuk.2020}. We run some experiments, which considered  a ring of particles similar to that described above, but now inclined with the angle, whose value was uniformly distributed in the  range $-0.01^\circ <i_2< 0.01^\circ $. We have found no qualitative differences with the results presented in Fig.\,\ref{fig-Particles}. 
Figure \ref{fig-Particlesinc} shows the result obtained for the inclinations of the particles after the $3 \times 10^6$\,yr of the tidal evolution of the system. We observe some excitation of the inclination in the domain surrounding Styx and Nix, that is expected during the passages through the strong 4/1 MMR. These results are in agreement with the orbital fits reported in \citet{Showalter+2015}.

The results obtained for the disc of the particles with the mass of $1.49\times 10^{15}$\,kg  are similar to those described above.   Thus, we can conclude that the simple model, which considers the accreting of the Charon's mass during the tidal expansion of the \PC binary, is effective to explain  the current locations of the small moons of Pluto,  without having to necessarily introduce the resonant transport mechanism.

\section{Conclusions}\label{sec.conclusions} 

We analyse the stability and dynamical evolution of the small moons in the PC system, namely, Styx, Nix, Kerberos and Hydra, during the tidal orbital expansion of Charon. We study the actual architecture of the system and also investigate how the tidal expansion of the PC binary system affects the orbital evolution of the small moons, extending previous studies.

The resonant structure around the PC binary exhibits $N/1$ and $N/2$ resonances, although most of them are weak in the low eccentricity and low inclination regimes. The 3/1 MMR is the most strong circumplanetary resonance even for low eccentricity and inclination of the moons (see Fig. \ref{fig1-conservative}). The current orbital configuration of the system does not show dynamical preferences of stable motion within the N/1 resonances, despite the four small moons are currently located close to the 3/1, 4/1, 5/1 and 6/1 MMRs. 

Using dynamical maps in the $a_{\s C}$--$e_{\s C}$ plane we reconstruct the tidal evolution of the \PC system 
with a small moon initially placed at its current position (see Fig.\,\ref{mapa-a1-e1}). We adopt a tidal model parameterized by the constant time lag, $\Delta t$. When the \PC system evolves assuming an initial value $e_{\s C} \sim 0.001$, the resonances are weaker than when the PC system evolves taking initial value $e_{\s C}$ $\sim 0.1$ or $0.2$. Thus, the evolution from initially quasi-circular orbits avoids chaotic regions associated to $N/1$ resonances, constraining the initial eccentricity of Charon. In addition, for small moons located near their current position, their semimajor axes do not secularly change, unless they are captured in a MMR due to the tidal expansion of the \PC binary.

The results of our numerical simulations indicate that it is possible to reproduce the current orbits of the four small moons of Pluto during the early tidal orbital expansion of Charon if we ignore the contribution of zonal harmonic $J_2$ and for a specific combination of initial conditions such as Charon’s eccentricity, semimajor axis and dissipation parameter $A$.
Hence, under the assumption of some specific values of physical parameters and initial semimajor axis and eccentricity of Charon, these results point out towards some specific scenarios (see Table \ref{tab-results}). For example, for Styx and Nix, an initial almost circular orbit of Charon is needed in order to obtain a final stable orbit of the small satellites. In addition,
a high dissipation in Pluto ($\Delta t_{\s P}$) allows for final orbital configurations of Styx in good agreement with their current orbits.

In the case of accreting Charon mass and considering an evolving $J_2$ with dissipation coefficient $A \gtrsim 14$, we find that the small moons can survive the rapid evolution of the Pluto-Charon system and the unlikely resonant transport of particles is no longer necessary. The ad-hoc mass evolution of Charon might allows the moons to cross some resonances and not to be trapped in specific positions of resonant motion. We also find that the eccentricity of Charon, $e_{\s C}$, when Charon crosses the 5/1 MMR with the actual position of Styx might be used to estimate the accretion time or dissipation coefficient in order to obtain survival moons as the system tidally evolves. Our numerical simulations considering a ring of particles between 34 $R_{\s C}$ and 60 $R_{\s C}$ show that a significant amount of particles remains in stable motion, with final orbits similar to the present small moons. 
Thus, the detection of additional small moons in future missions to the Pluto system should not be ruled out.

Further work is necessary to constrain the mass accretion in a more self consistent scenario \citep[e.g.,][]{Kokubo+2000, Kokubo+1998}. In addition, our model could be improved by considering mutual perturbations between the small moons \citep{Showalter+2015, lee+peale2006}; by using a more realistic model for tidal dissipation within icy bodies (see \citealt{ferraz2013,folonier2018}) or by considering the obliquity of Charon and perturbations from the Sun and Neptune \citep[see e.g.,][]{Correia+2020}.

\begin{acknowledgements}
$N$-body computations were performed at Mulatona Cluster from CCAD - UNC, which is part of SNCAD - MinCyT, Argentina. 
The construction of the dynamical maps has made use of the facilities of the Laboratory of Astroinformatics (IAG/USP, NAT/Unicsul), whose purchase was made possible by FAPESP (grant 2009/54006-4) and the INCT-A. TAM was supported by the Brazilian agencies Conselho Nacional de Desenvolvimento Cient\'ifico e Tecnol\'ogico (CNPq), and S\~ao Paulo Research Foundation (FAPESP, grant 2016/13750-6). 
{We acknowledge the program Professor Visitante do Exterior (PVE)
 from Pró-Reitoria de Pós-Graduação - Universidade de São Paulo. Edital PRPG 02/2019.}

\end{acknowledgements}

\bibliographystyle{aa}

\bibliography{pluto} 

\begin{thebibliography}{47}
\expandafter\ifx\csname natexlab\endcsname\relax\def\natexlab#1{#1}\fi

\bibitem[{{Asphaug} {et~al.}(2006){Asphaug}, {Agnor}, \&
  {Williams}}]{asphaug2006}
{Asphaug}, E., {Agnor}, C.~B., \& {Williams}, Q. 2006, \nat, 439, 155

\bibitem[{{Bromley} \& {Kenyon}(2020)}]{Bromley+2020arXiv200613901B}
{Bromley}, B.~C. \& {Kenyon}, S.~J. 2020, arXiv e-prints, arXiv:2006.13901

\bibitem[{{Brozovi{\'c}} {et~al.}(2015){Brozovi{\'c}}, {Showalter}, {Jacobson},
  \& {Buie}}]{Brozovic+2015}
{Brozovi{\'c}}, M., {Showalter}, M.~R., {Jacobson}, R.~A., \& {Buie}, M.~W.
  2015, \icarus, 246, 317

\bibitem[{{Buie} {et~al.}(2012){Buie}, {Tholen}, \& {Grundy}}]{Buie.2012}
{Buie}, M.~W., {Tholen}, D.~J., \& {Grundy}, W.~M. 2012, \aj, 144, 15

\bibitem[{{Canup}(2005)}]{Canup2005}
{Canup}, R.~M. 2005, Science, 307, 546

\bibitem[{{Canup}(2011)}]{Canup+2011}
{Canup}, R.~M. 2011, \aj, 141, 35

\bibitem[{{Cheng} {et~al.}(2014{\natexlab{a}}){Cheng}, {Lee}, \&
  {Peale}}]{Chen+2014a}
{Cheng}, W.~H., {Lee}, M.~H., \& {Peale}, S.~J. 2014{\natexlab{a}}, \icarus,
  233, 242

\bibitem[{{Cheng} {et~al.}(2014{\natexlab{b}}){Cheng}, {Peale}, \&
  {Lee}}]{Chen+2014b}
{Cheng}, W.~H., {Peale}, S.~J., \& {Lee}, M.~H. 2014{\natexlab{b}}, \icarus,
  241, 180

\bibitem[{{Cincotta} \& {Sim{\'o}}(2000)}]{Cincotta.Simo.1999}
{Cincotta}, P.~M. \& {Sim{\'o}}, C. 2000, \aaps, 147, 205

\bibitem[{{Correia}(2020)}]{Correia+2020}
{Correia}, A. C.~M. 2020, \aap, 644, A94

\bibitem[{{Cuello} \& {Giuppone}(2019)}]{Cuello&Giuppone2019}
{Cuello}, N. \& {Giuppone}, C.~A. 2019, \aap, 628, A119

\bibitem[{{{\'C}uk} {et~al.}(2020){{\'C}uk}, {El Moutamid}, \&
  {Tiscareno}}]{Cuk.2020}
{{\'C}uk}, M., {El Moutamid}, M., \& {Tiscareno}, M.~S. 2020, The Planetary
  Science Journal, 1, 22

\bibitem[{{Desch}(2015)}]{Desch+2015}
{Desch}, S.~J. 2015, \icarus, 246, 37

\bibitem[{{Dvorak} {et~al.}(2004){Dvorak}, {Pilat-Lohinger}, {Schwarz}, \&
  {Freistetter}}]{Dvorak.etal.2004}
{Dvorak}, R., {Pilat-Lohinger}, E., {Schwarz}, R., \& {Freistetter}, F. 2004,
  \aap, 426, L37

\bibitem[{{Farinella} {et~al.}(1979){Farinella}, {Milani}, {Nobili}, \&
  {Valsecchi}}]{Farinella+1979}
{Farinella}, P., {Milani}, A., {Nobili}, A.~M., \& {Valsecchi}, G.~B. 1979,
  Moon and Planets, 20, 415

\bibitem[{{Ferraz-Mello}(2013)}]{ferraz2013}
{Ferraz-Mello}, S. 2013, Celestial Mechanics and Dynamical Astronomy, 116, 109

\bibitem[{{Folonier} {et~al.}(2018){Folonier}, {Ferraz-Mello}, \&
  {Andrade-Ines}}]{folonier2018}
{Folonier}, H.~A., {Ferraz-Mello}, S., \& {Andrade-Ines}, E. 2018, Celestial
  Mechanics and Dynamical Astronomy, 130, 78

\bibitem[{{Gallardo} {et~al.}(2021){Gallardo}, {Beaug{\'e}}, \&
  {Giuppone}}]{Gallardo+2021}
{Gallardo}, T., {Beaug{\'e}}, C., \& {Giuppone}, C.~A. 2021, \aap, 646, A148

\bibitem[{{Holman} \& {Wiegert}(1999)}]{Holman+Wiegert1999}
{Holman}, M.~J. \& {Wiegert}, P.~A. 1999, \aj, 117, 621

\bibitem[{{Kenyon} \& {Bromley}(2019{\natexlab{a}})}]{Kenyon+2019b}
{Kenyon}, S.~J. \& {Bromley}, B.~C. 2019{\natexlab{a}}, \aj, 158, 69

\bibitem[{{Kenyon} \& {Bromley}(2019{\natexlab{b}})}]{Kenyon+BromleyIII}
{Kenyon}, S.~J. \& {Bromley}, B.~C. 2019{\natexlab{b}}, \aj, 158, 142

\bibitem[{{Kenyon} \& {Bromley}(2019{\natexlab{c}})}]{Kenyon+2019a}
{Kenyon}, S.~J. \& {Bromley}, B.~C. 2019{\natexlab{c}}, \aj, 157, 79

\bibitem[{{Kenyon} \& {Bromley}(2021)}]{Kenyon+2021}
{Kenyon}, S.~J. \& {Bromley}, B.~C. 2021, arXiv e-prints, arXiv:2102.11311

\bibitem[{{Kokubo} \& {Ida}(1998)}]{Kokubo+1998}
{Kokubo}, E. \& {Ida}, S. 1998, \icarus, 131, 171

\bibitem[{{Kokubo} {et~al.}(2000){Kokubo}, {Ida}, \& {Makino}}]{Kokubo+2000}
{Kokubo}, E., {Ida}, S., \& {Makino}, J. 2000, \icarus, 148, 419

\bibitem[{{Lainey}(2016)}]{Lainey2016}
{Lainey}, V. 2016, Celestial Mechanics and Dynamical Astronomy, 126, 145

\bibitem[{{Lee} \& {Peale}(2006)}]{lee+peale2006}
{Lee}, M.~H. \& {Peale}, S.~J. 2006, \icarus, 184, 573

\bibitem[{{Lithwick} \& {Wu}(2008{\natexlab{a}})}]{Lithwick+Wu2008}
{Lithwick}, Y. \& {Wu}, Y. 2008{\natexlab{a}}, arXiv e-prints, arXiv:0802.2951

\bibitem[{{Lithwick} \& {Wu}(2008{\natexlab{b}})}]{Lithwick+2008}
{Lithwick}, Y. \& {Wu}, Y. 2008{\natexlab{b}}, arXiv e-prints, arXiv:0802.2939

\bibitem[{{McKinnon} {et~al.}(2017){McKinnon}, {Stern}, {Weaver}, {Nimmo},
  {Bierson}, {Grundy}, {Cook}, {Cruikshank}, {Parker}, {Moore}, {Spencer},
  {Young}, {Olkin}, {Ennico Smith}, {New Horizons Geology}, {Imaging}, \&
  {Composition Theme Teams}}]{McKinnon+2017}
{McKinnon}, W.~B., {Stern}, S.~A., {Weaver}, H.~A., {et~al.} 2017, \icarus,
  287, 2

\bibitem[{{Michtchenko} \& {Rodr{\'\i}guez}(2011)}]{michtchenko+rodriguez2011}
{Michtchenko}, T.~A. \& {Rodr{\'\i}guez}, A. 2011, \mnras, 415, 2275

\bibitem[{{Mignard}(1979)}]{1979M&P....20..301M}
{Mignard}, F. 1979, Moon and Planets, 20, 301

\bibitem[{{Pires Dos Santos} {et~al.}(2011){Pires Dos Santos}, {Giuliatti
  Winter}, \& {Sfair}}]{Pires+2011}
{Pires Dos Santos}, P.~M., {Giuliatti Winter}, S.~M., \& {Sfair}, R. 2011,
  \mnras, 410, 273

\bibitem[{{Pires dos Santos} {et~al.}(2012){Pires dos Santos}, {Morbidelli}, \&
  {Nesvorn{\'y}}}]{Pires+2012}
{Pires dos Santos}, P.~M., {Morbidelli}, A., \& {Nesvorn{\'y}}, D. 2012,
  Celestial Mechanics and Dynamical Astronomy, 114, 341

\bibitem[{{Quillen} {et~al.}(2017){Quillen}, {Nichols-Fleming}, {Chen}, \&
  {Noyelles}}]{Quillen+2017}
{Quillen}, A.~C., {Nichols-Fleming}, F., {Chen}, Y.-Y., \& {Noyelles}, B. 2017,
  \icarus, 293, 94

\bibitem[{{Ramos} {et~al.}(2015){Ramos}, {Correa-Otto}, \&
  {Beaug{\'e}}}]{Ramos.etal.2015}
{Ramos}, X.~S., {Correa-Otto}, J.~A., \& {Beaug{\'e}}, C. 2015, Celestial
  Mechanics and Dynamical Astronomy, 123, 453

\bibitem[{{Rodr{\'\i}guez} {et~al.}(2013){Rodr{\'\i}guez}, {Giuppone}, \&
  {Michtchenko}}]{Rodriguez+2013}
{Rodr{\'\i}guez}, A., {Giuppone}, C.~A., \& {Michtchenko}, T.~A. 2013,
  Celestial Mechanics and Dynamical Astronomy, 117, 59

\bibitem[{{Showalter} \& {Hamilton}(2015)}]{Showalter+2015}
{Showalter}, M.~R. \& {Hamilton}, D.~P. 2015, \nat, 522, 45

\bibitem[{{Smullen} \& {Kratter}(2017)}]{Smullen+2017}
{Smullen}, R.~A. \& {Kratter}, K.~M. 2017, \mnras, 466, 4480

\bibitem[{{Tholen} {et~al.}(2008){Tholen}, {Buie}, {Grundy}, \&
  {Elliott}}]{Tholen+2008}
{Tholen}, D.~J., {Buie}, M.~W., {Grundy}, W.~M., \& {Elliott}, G.~T. 2008, \aj,
  135, 777

\bibitem[{{Walsh} \& {Levison}(2015)}]{Walsh+2015}
{Walsh}, K.~J. \& {Levison}, H.~F. 2015, \aj, 150, 11

\bibitem[{{Ward} \& {Canup}(2006)}]{Ward+2006}
{Ward}, W.~R. \& {Canup}, R.~M. 2006, Science, 313, 1107

\bibitem[{{Weaver} {et~al.}(2016){Weaver}, {Buie}, {Buratti}, {Grundy},
  {Lauer}, {Olkin}, {Parker}, {Porter}, {Showalter}, {Spencer}, {Stern},
  {Verbiscer}, {McKinnon}, {Moore}, {Robbins}, {Schenk}, {Singer}, {Barnouin},
  {Cheng}, {Ernst}, {Lisse}, {Jennings}, {Lunsford}, {Reuter}, {Hamilton},
  {Kaufmann}, {Ennico}, {Young}, {Beyer}, {Binzel}, {Bray}, {Chaikin}, {Cook},
  {Cruikshank}, {Dalle Ore}, {Earle}, {Gladstone}, {Howett}, {Linscott},
  {Nimmo}, {Parker}, {Philippe}, {Protopapa}, {Reitsema}, {Schmitt}, {Stryk},
  {Summers}, {Tsang}, {Throop}, {White}, \& {Zangari}}]{Weaver+2016}
{Weaver}, H.~A., {Buie}, M.~W., {Buratti}, B.~J., {et~al.} 2016, Science, 351,
  aae0030

\bibitem[{{Weaver} {et~al.}(2006){Weaver}, {Stern}, {Mutchler}, {Steffl},
  {Buie}, {Merline}, {Spencer}, {Young}, \& {Young}}]{Weaver+2006}
{Weaver}, H.~A., {Stern}, S.~A., {Mutchler}, M.~J., {et~al.} 2006, \nat, 439,
  943

\bibitem[{{Woo} \& {Lee}(2018)}]{Woo+2018}
{Woo}, J. M.~Y. \& {Lee}, M.~H. 2018, \aj, 155, 175

\bibitem[{{Woo} \& {Lee}(2020)}]{Woo+2020}
{Woo}, J. M.~Y. \& {Lee}, M.~H. 2020, \aj, 159, 277

\bibitem[{{Youdin} {et~al.}(2012){Youdin}, {Kratter}, \&
  {Kenyon}}]{Youdin+2012}
{Youdin}, A.~N., {Kratter}, K.~M., \& {Kenyon}, S.~J. 2012, \apj, 755, 17

\end{thebibliography}
\appendix

\section{Dynamical evolution of small moons}\label{Append-0}
\begin{figure*}[h!]
  \centering
\mbox{\includegraphics[width=0.7\textwidth]{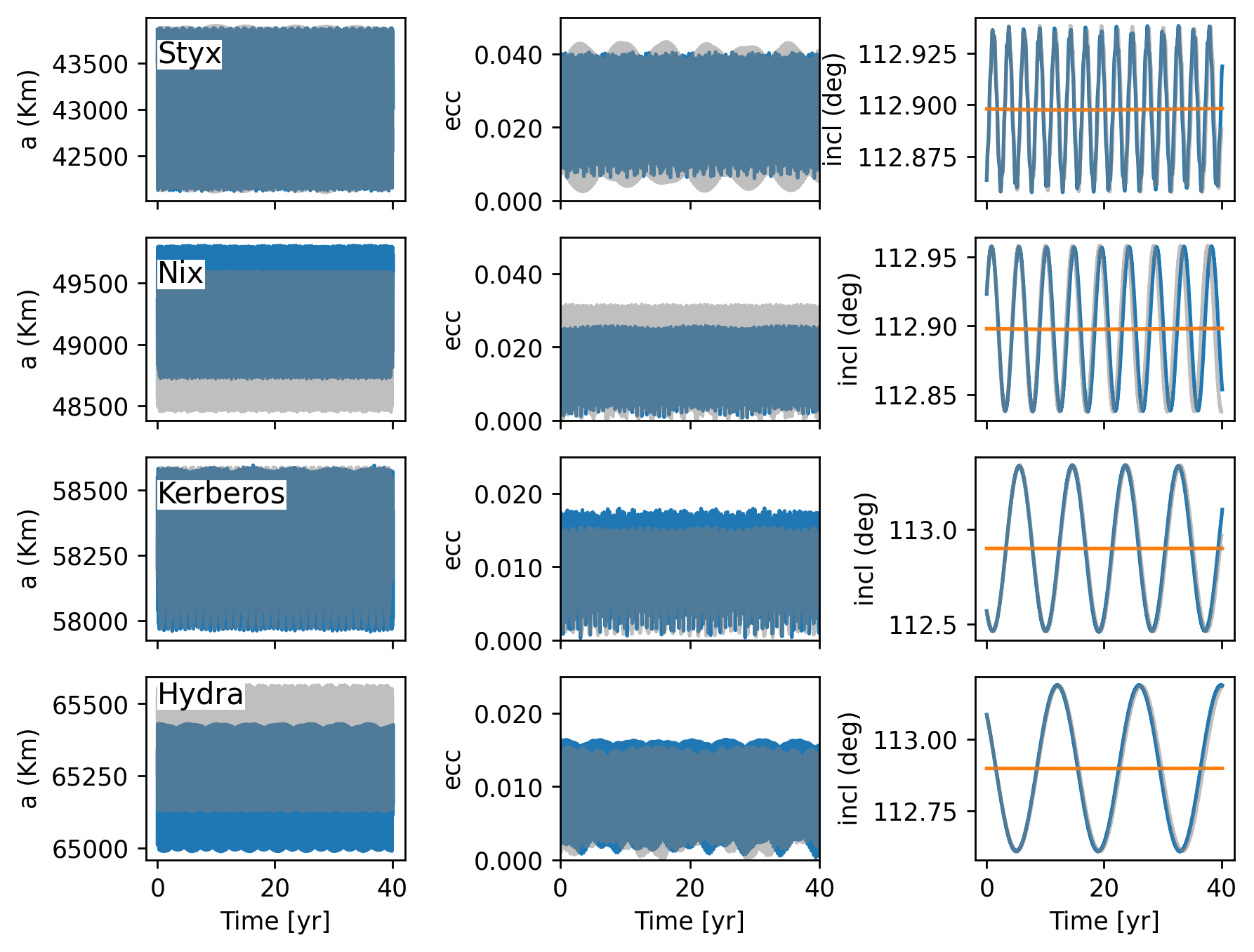}} \\
\caption{Orbital evolution Styx, Nix, Hydra, and Kerberos around the PC binary, from top to bottom row, respectively. We show the evolution of semimajor axis (left panels), eccentricity (middle panels) and inclination (right panels) for 40 yr. Blue lines correspond to data from JPL, gray lines with our N-body integrations. Orange lines corresponds to integrated inclination of Charon. }
  \label{fig-evol}
  \vspace{-1em}
  \end{figure*}

Here we provide complementary information to construct the Table \ref{tab.Pluto}. In Figure \ref{fig-evol} we show N-body integration of three-body problem for a time-span of 40 yr (gray lines), together with ephemerides taken from the \textit{JPL-Horizons}.  
Amplitude oscillation of the orbital elements are resumed in our Table \ref{tab.Pluto}, and agree with the colour scale presented in Fig \ref{fig1-conservative}. 
It is interesting to look first at JPL ephemerides (blue lines in the figure). Left panels indicate the evolution of the semimajor axis that presents amplitude of oscillations from top to bottom of 1761 km, 1086 km, 637 km, and 443 km (roughly $\Delta a=1.5 R_P, 0.9 R_P, 0.5 R_P$, and $0.4 R_P$) for Styx, Nix, Kerberos, and Hydra, respectively. We remind that orbital elements are given with respect to the barycenter of \PC system. Osculating eccentricities present higher variations for closer moons and with lower amplitude for exterior moons (being in order, 0.034, 0.025, 0.018, 0.016). These variations are of the same order than those variations presented in Fig 2 in \citet{Woo+2018} and Fig 1 in \cite{Woo+2020}, although the authors used eccentricity for Charon from JPL solution PLU043 which is older and an order of magnitude lower than the current determinations (solution PLU058). Finally, inclination evolution (right panels in Figure \ref{fig-evol}) show oscillations around the midplane of Pluto-Charon binary (denoted by orange line and which is nearly coincident between N-body integrations and JPL query data). It is evident that the small moons can be either above or below the plane with different phases, and this inspires us in sections \ref{sec.constantmass} and \ref{sec.massgrowth} to choose the \PC orbit as reference plane, thus for example defining inclinations randomly chosen around zero degrees. 

Our N-body integrations (gray lines in Fig.\ref{fig-evol}) shown almost the same amplitudes than JPL query (see also Table \ref{tab.Pluto}) but there is an evident shift in the evolution of semimajor axis of Nix and Hydra, although $\Delta a$ is almost the same. We believe that the differences in the evolution of semimajor axis in our simulations and JPL data could be originated in the more sophisticated JPL-model that considers a significant number of perturbations, including the Sun and Neptune. 

\begin{figure*}
  \centering
\mbox{\includegraphics[width=0.9\textwidth]{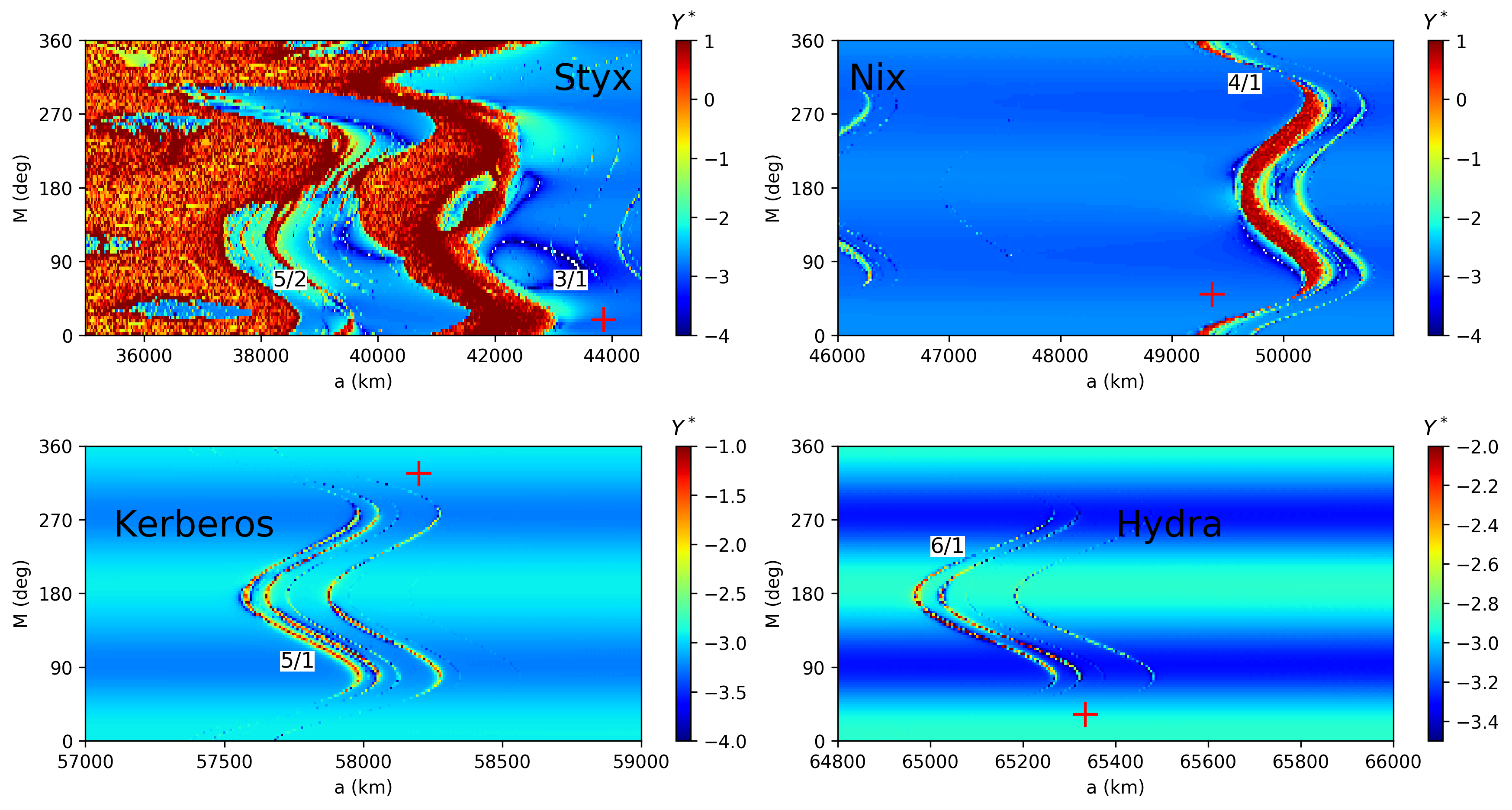}} 
\caption{Stability maps on the $a$--$M$ plane using Megno colour scale for the four small moons. All other orbital parameters set to JPL from Table \ref{tab.Pluto}. The current positions of the moons are shown by red crosses. The locations of the main MMRs are indicated. Small deviations from initial conditions within the estimated errors from PLU043 can put the moons inside the MMR regime.}
  \label{fig-mapaM}
  \vspace{-1em}
  \end{figure*}

Figure \ref{fig-mapaM} shows four dynamical maps in the plane semimajor axis and mean anomaly, constructed using the Megno colour scale \citep[e.g., Megno=$\langle Y \rangle$, Mean Exponential Growth of Nearby Orbits,][]{Cincotta.Simo.1999}, ($Y^* = log_{10}(\langle Y \rangle -2))$. Red crosses indicate the positions of each moon at 2021/1/1 and all other parameters in the plane correspond to those retrieved from JPL (see Table \ref{tab.Pluto}). Megno allow us to identify the separatrices of each resonance as chaotic (reddish colour of each panel). The MMR 3/1 is evidently more asymmetric than the 4/1, 5/1, and 6/1 MMRs because of its proximity to the Pluto-Charon binary. For example, for Styx plane the separatrices show that the initial mean anomaly modifies the location of the resonance in $\sim$1000 kms. Also, the MMR 3/1 and 4/1 are wider than 5/1 and 6/1 MMR because the moons experiment greater excursions in eccentricity. These maps also helps to understand that stability maps in planes $a-e$ strongly depend on the chosen value of mean anomaly (in our Figure \ref{fig1-conservative} correspond to those from Styx, $M_S=18.79^\circ$).  

\newpage
\section{Survival times of the disc described by simplified models}\label{Append-1}

Here we present a complementary information to that discussed in Sect.\,\ref{sec.constantmass}. Now we investigate the survival times of the disc of small moons applying the model, which uses fixed values of the zonal harmonic of the Pluto's oblateness, either $J_2=0$ or $J_2= 6 \times 10^{-5}$ (this last value was determined from the shape of Pluto in \citep[][]{Correia+2020}). We remember that our model consists of the tidally evolving \PC binary and a disc of small moons extending from $10 {R_{\s P}}$ to $55 {R_{\s P}}$ (for more details, see Sect.\,\ref{sec.constantmass}).

\begin{figure}
  \centering
\mbox{\includegraphics[width=0.9\columnwidth]{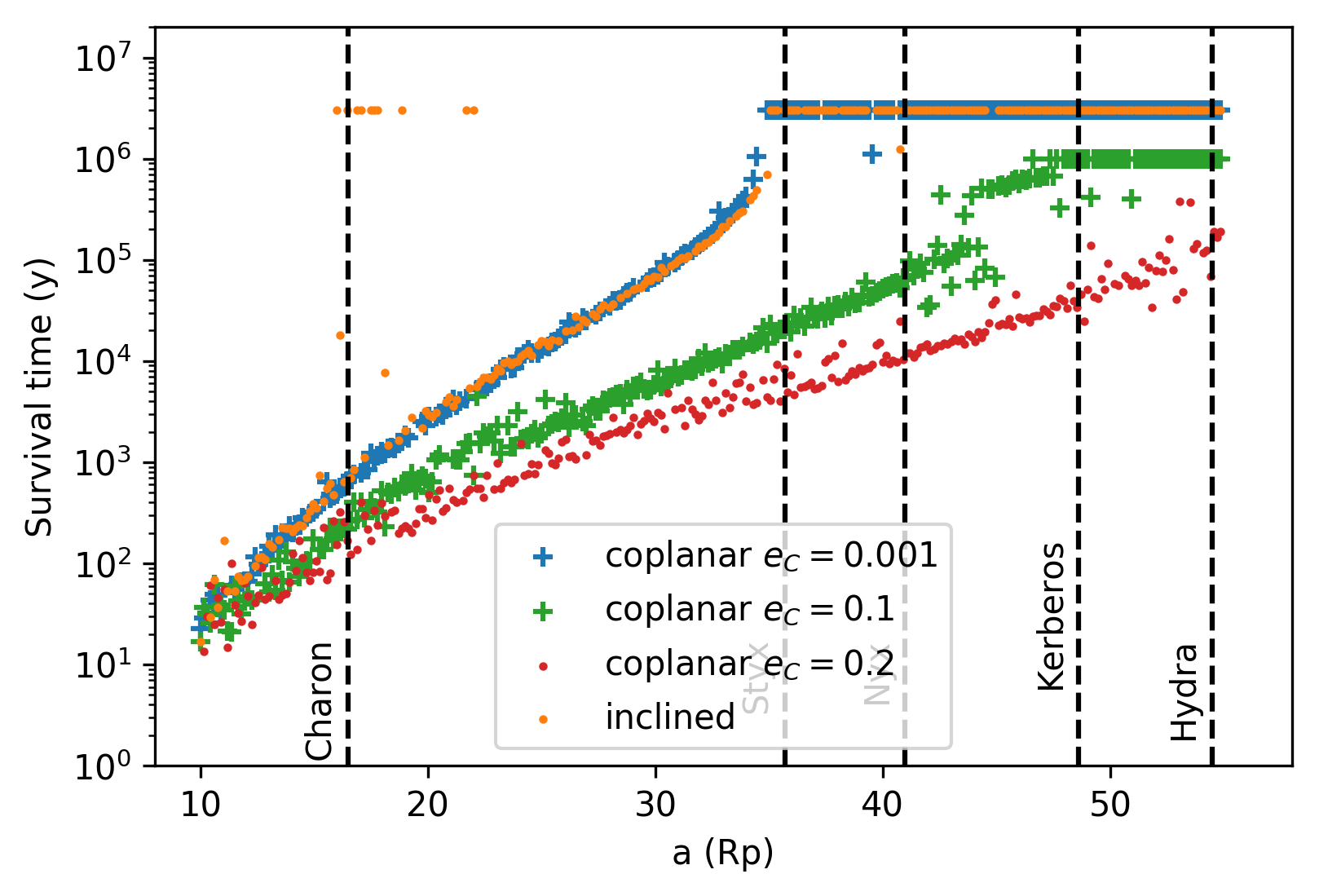}} \\
\caption{Survival times of the test particles from different discs, all extending from $10 {R_{\s P}} <a < 55 {R_{\s P}}$, for the dissipation parameter $A=10$. We use different symbols/colours for initially coplanar and inclined discs, as well as for different initial eccentricities of  the Charon's orbit. {We do not consider $J_2$ effects, and} the Charon mass is conserved during the integrations. (see text for more details).}
  \label{fig-survival}
  \vspace{-1em}
  \end{figure}

Figure \ref{fig-survival} shows the case when the effect of the zonal harmonic is not considered ($J_2 = 0$). For Charon starting on the nearly circular orbit, the survival times of the test moons on the planar orbits (blue crosses) monotonously increase (in log scale) {with the increasing semimajor axis}, reaching $10^3$\,yr for particles in the internal disk ($10 R_P < a< 20 R_P$). The particles at the Styx's position and beyond it ($a> 35 R_P$) show the survival times greater than $3 \times 10^6$\,yr. Moreover, the numerical integrations of the individual orbits show that the moon's motions are stable for more than $1 \times 10^8$ orbital periods of the \PC binary.

\begin{figure}
  \centering
\mbox{\includegraphics[width=0.8\columnwidth]{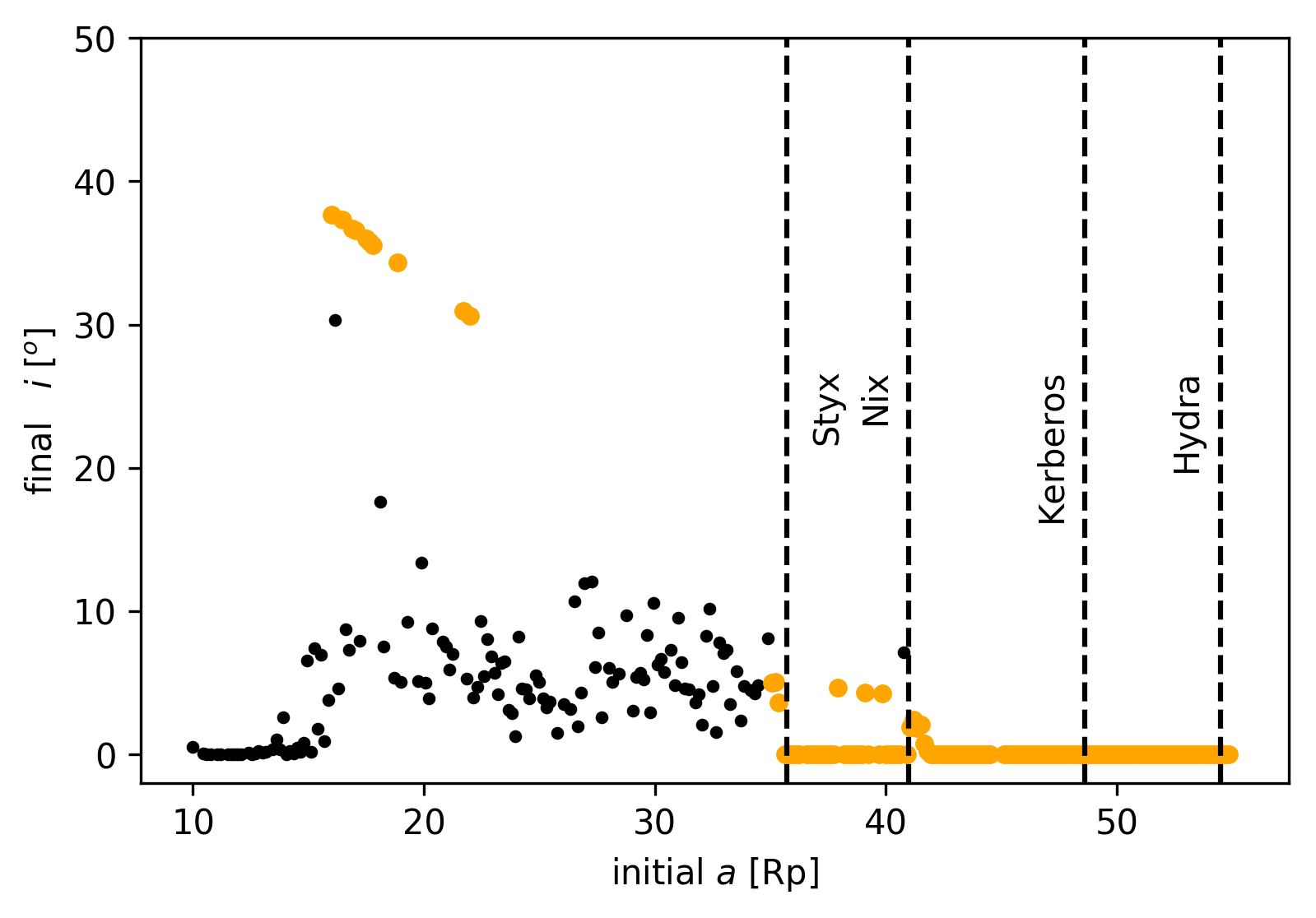}} \\
\caption{Distribution of the final inclinations of a thin disc of particles shown in Figure \ref{fig-survival}.  The particles ejected from the \PC system are identified by black dots. }
  \label{fig-survivalinc}
  \vspace{-1em}
  \end{figure}

When we consider a thin disk of small moons with inclinations uniformly distributed between $-0.01^\circ$ and $0.01^\circ$ (Figure \ref{fig-survival}, orange dots), we observe that survival times are  similar to those obtained for a coplanar disk. A difference is observed essentially for the moons initially located at the Charon's position. We analyse the final inclinations of the particles in the thin disk in Fig.\,\ref{fig-survivalinc} and observe that the orbits of the surviving moons around 20 $R_P$ reach inclinations up to $25^\circ$. The inclinations of the particles located inside the limit given by the HWC, are dispersed around the mean value of $5^\circ$. Also, some portion of the test particles near the Nix position is excited in inclination up to $5^\circ$, but survive during the timespan of simulations.

Figure \ref{fig-survival} also presents the survival times obtained for the coplanar disk, but with higher initial values of Charon eccentricities, namely, $e_C=0.1$ and $e_C=0.2$ (green crosses and red dots, respectively). These values put constraints on the initial rotational periods of Pluto, being now $P_0=0.13791$\,d and $P_0=0.13593$\,d, respectively. For $e_C=0.1$,  only particles beyond the Kerberos orbit survive during $10^6$\,yr. Setting the initial values of  $e_C=0.2$ causes a faster depletion of the initial disc, at $<10^5$\,yr. We do not find differences choosing initially aligned or anti-aligned longitudes of the particle's pericenter and the pericenter of the Charon's orbit.

{The Figure \ref{fig-survival} can be directly compared with Fig. 10 in \citet{Woo+2018}, where almost all the particles in the range $a>30 R_P$ survived. However, their model was constructed for initial $e_C=0$, massless particles, and $A=40$. We can observe in our simulations that moderate values ($A=10$) allow the survival of particles. However, \citet{Woo+2018} do not explore cases in almost initially circular Charon ($e_C\sim0$), neither inclined particles. }

\begin{figure}
  \centering
\mbox{\includegraphics[width=0.9\columnwidth]{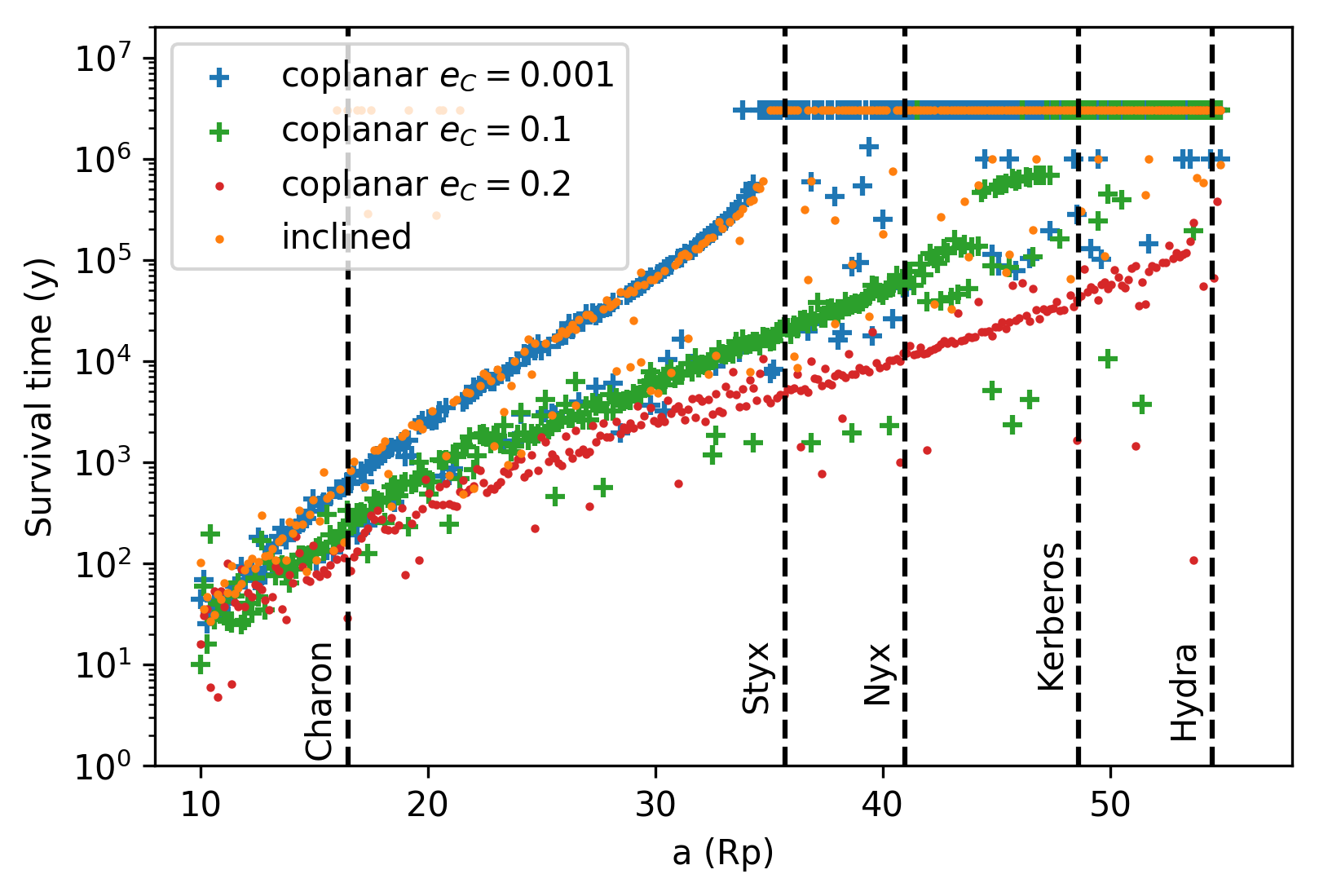}} \\
\caption{Same as in Fig.\,\ref{fig-survival}, except considering the effect of {fixed} zonal harmonic $J_2=6\times 10^{-5}$. }
  \label{fig-survival-J2fix}
  \vspace{-1em}
  \end{figure}

 The results obtained setting a fixed value of $J_2$ are present in Fig.\,\ref{fig-survival-J2fix}. We remember that the static value of $J_2=6\times 10^{-5}$ were derived  from the shape of Pluto in \citet{Correia+2020}. We can observe that, for all initial eccentricities of Charon, the particles evolves almost in the same way as shown in Fig.\,\ref{fig-survival}.

\end{document}